\documentclass[12pt]{article}
\usepackage{amsmath,amssymb,amsthm,amsxtra,overpic,bbm,bm,epsfig,subfigure}
\usepackage{mathrsfs}
\usepackage{graphicx}
\usepackage{color}
\usepackage{comment}
\usepackage{epstopdf}
\usepackage{float}
\usepackage{slashed,stmaryrd}

\def\thefootnote{\fnsymbol{footnote}}

\begin{document}

\vspace{0.2cm}

\begin{center}
{\Large\bf One-loop radiative correction to the Toshev relation \\
for neutrino oscillations in matter}
\end{center}

\vspace{0.2cm}

\begin{center}
{\bf Zhi-zhong Xing~$^{a,b,c}$},
\quad
{\bf Jun-yu Zhu~$^{a,b}$} \footnote{E-mail: zhujunyu22@mails.ucas.ac.cn (corresponding author)}
\\
{\small $^a$Institute of High Energy Physics, Chinese Academy of
Sciences, Beijing 100049, China \\
$^b$School of Physical Sciences, University of Chinese Academy of Sciences,
Beijing 100049, China \\
$^c$Center for High Energy Physics, Peking University, Beijing 100871, China}
\end{center}

\vspace{1.5cm}

\begin{abstract}
The one-loop electroweak radiative corrections to coherent forward neutrino
scattering in a medium slightly violates the tree-level universality of
neutral-current contributions of three neutrino flavors to the matter potential
that is relevant to neutrino oscillations in matter. We examine this small but
nontrivial quantum effect by deriving the differential equations
of those effective neutrino oscillation quantities with respect to the electron
number density of matter. The tree-level Toshev relation $\sin 2 \tilde{\theta}^{}_{23}
\sin\tilde{\delta} = \sin 2\theta^{}_{23} \sin\delta$, which links the fundamental
flavor-mixing and CP-violating parameters $\big(\theta^{}_{23}, \delta\big)$ to their
matter-corrected counterparts $\big(\tilde{\theta}^{}_{23}, \tilde{\delta}\big)$ in the
standard parametrization of the $3\times 3$ lepton flavor mixing matrix, is shown to
be modified at the one-loop level. We numerically illustrate the significance of
such effects for neutrino and antineutrino oscillations in dense matter.
\end{abstract}

\begin{flushleft}
\hspace{0.8cm} PACS number(s): 14.60.Pq, 25.30.Pt
\end{flushleft}

\def\thefootnote{\arabic{footnote}}
\setcounter{footnote}{0}

\newpage

\section{Introduction}

When a neutrino beam travels in an electrically neutral and unpolarized medium,
the coherent forward neutrino scattering with electrons via weak charged-current
(CC) interactions and with electrons, protons and neutrons via weak
neutral-current (NC) interactions can definitely modify the behaviors of neutrino
oscillations --- a striking phenomenon which is usually referred to as the
Mikheev-Smirnov-Wolfenstein (MSW) matter
effect~\cite{Wolfenstein:1977ue,Mikheev:1986gs,Mikheev:1986wj}. The
effective Hamiltonian responsible for the evolution of three active neutrino
flavors in matter is composed of the vacuum term and the matter potential term
as follows~\cite{Kuo:1989qe,Xing:2003ez}:
\begin{eqnarray}
{\cal H}^{}_{\rm eff} = \frac{1}{2 E} \hspace{0.07cm}
U \left(\begin{matrix} m^2_1 & 0 & 0 \cr 0
& m^2_2 & 0 \cr 0 & 0 & m^2_3 \end{matrix}\right) U^\dagger +
\left(\begin{matrix} {\cal V}^{}_e & 0 & 0 \cr
0 & {\cal V}^{}_\mu & 0 \cr 0 & 0 & {\cal V}^{}_\tau \end{matrix}\right)
\equiv \frac{1}{2 E} \hspace{0.07cm} V \left(\begin{matrix}
\widetilde{m}^{2}_1 & 0 & 0 \cr 0 & \widetilde{m}^2_2 & 0 \cr 0 & 0 &
\widetilde{m}^2_3 \end{matrix}\right) V^\dagger \; ,
\label{eq:H}
%     (1)
\end{eqnarray}
where $m^{}_i$ (for $i=1,2,3$) and $U$ stand respectively for the neutrino masses
and the Pontecorvo-Maki-Nakagawa-Skata (PMNS) flavor mixing
matrix~\cite{Pontecorvo:1957cp,Maki:1962mu,Pontecorvo:1967fh} in vacuum,
$\widetilde{m}^{}_i$ (for $i=1,2,3$) and $V$ are defined to be their
effective counterparts in matter, and ${\cal V}^{}_\alpha$ (for $\alpha = e, \mu, \tau$)
describe the effects of coherent forward neutrino scattering with matter. At
the tree level we have ${\cal V}^{}_e = {\cal V}^{}_{\rm CC} + {\cal V}^{}_{\rm NC}$
and ${\cal V}^{}_\mu = {\cal V}^{}_\tau = {\cal V}^{}_{\rm NC}$, where
${\cal V}^{}_{\rm CC} = \sqrt{2} \hspace{0.04cm} G^{}_{\rm F} N^{}_e$
and ${\cal V}^{}_{\rm NC} = -G^{}_{\rm F}/\sqrt{2}
\left[\left(1 - 4\sin^2\theta^{}_{\rm w} \right) \left(N^{}_e - N^{}_p\right)
+ N^{}_n\right]$ with $G^{}_{\rm F}$ being the Fermi coupling constant,
$\theta^{}_{\rm w}$ being the Weinberg angle of weak interactions, and
$N^{}_e$, $N^{}_p$ and $N^{}_n$ being the number densities of electrons, protons
and neutrons in matter. Given the fact that $N^{}_e = N^{}_p$ holds for a normal
medium, one is simply left with ${\cal V}^{}_{\rm NC} = - G^{}_{\rm F} N^{}_n/\sqrt{2}$.
Since the tree-level result of ${\cal V}^{}_{\rm NC}$ is universal for $\nu^{}_e$,
$\nu^{}_\mu$ and $\nu^{}_\tau$ neutrinos, it has no physical impact on neutrino
oscillations and hence can be ignored in the standard three-flavor scheme.

The $3\times 3$ PMNS lepton flavor mixing matrix $U$ in Eq.~(\ref{eq:H}) can be
parametrized in a way advocated by the Particle Data Group~\cite{Zyla:2020zbs}:
$U = P^{}_l O^{}_{23} O^{}_\delta O^{}_{13} O^\dagger_\delta O^{}_{12} P^{}_\nu$;
namely,
\begin{eqnarray}
U & = & P^{}_l
\left(\begin{matrix}
1 & 0 & 0 \cr 0 & c^{}_{23} & s^{}_{23} \cr 0 & -s^{}_{23} &  c^{}_{23} \cr
\end{matrix} \right)
\left(\begin{matrix}
1 & 0 & 0 \cr 0 & 1 & 0 \cr 0 & 0 & e^{+{\rm i} \delta} \cr
\end{matrix} \right)
\left(\begin{matrix}
c^{}_{13} & 0 & s^{}_{13} \cr 0 & 1 & 0 \cr
-s^{}_{13} & 0 & c^{}_{13} \cr \end{matrix} \right)
\left(\begin{matrix}
1 & 0 & 0 \cr 0 & 1 & 0 \cr 0 & 0 & e^{-{\rm i} \delta} \cr
\end{matrix} \right)
\left(\begin{matrix}
c^{}_{12} & s^{}_{12} & 0 \cr -s^{}_{12} & c^{}_{12} & 0 \cr
0 & 0 & 1 \cr \end{matrix} \right) P^{}_\nu \hspace{0.4cm}
\nonumber \\
& = & P^{}_l
\left(\begin{matrix}
c^{}_{12} c^{}_{13} & s^{}_{12} c^{}_{13} &
s^{}_{13} e^{-{\rm i} \delta} \cr
-s^{}_{12} c^{}_{23} - c^{}_{12}
s^{}_{13} s^{}_{23} e^{{\rm i} \delta} & c^{}_{12} c^{}_{23} -
s^{}_{12} s^{}_{13} s^{}_{23} e^{{\rm i} \delta} & c^{}_{13}
s^{}_{23} \cr
s^{}_{12} s^{}_{23} - c^{}_{12} s^{}_{13} c^{}_{23}
e^{{\rm i} \delta} &- c^{}_{12} s^{}_{23} - s^{}_{12} s^{}_{13}
c^{}_{23} e^{{\rm i} \delta} &  c^{}_{13} c^{}_{23} \cr
\end{matrix} \right) P^{}_\nu \; ,
\label{eq:U}
%     (2)
\end{eqnarray}
where $c^{}_{ij} \equiv \cos\theta^{}_{ij}$ and $s^{}_{ij} \equiv \sin\theta^{}_{ij}$
(for $ij = 12, 13, 23$) with $\theta^{}_{ij}$ being the flavor mixing angles, $\delta$
denotes the Dirac phase responsible for CP violation in neutrino oscillations, $P^{}_l$
is a diagonal phase matrix which is sensitive to rephasing the charged-lepton fields
but has no physical significance, and $P^{}_\nu$ is another diagonal
phase matrix relevant to Majorana neutrinos but has nothing to do with neutrino
oscillations. We adopt the same parametrization for the effective PMNS matrix
$V$ in matter, whose Dirac phase and flavor mixing angles are denoted respectively
as $\tilde{\delta}$ and $\tilde{\theta}^{}_{ij}$ (for $ij = 12, 13, 23$). Note
that ${\cal V}^{}_\mu = {\cal V}^{}_\tau$ allows $O^{}_{23}$, $O^{}_\delta$ and
$O^\dagger_\delta$ to commute with the diagonal matter potential term in
Eq.~({\ref{eq:H}), and this property leads us to the intriguing Toshev
relation~\cite{Toshev:1991ku,Freund:2001pn,Zhou:2011xm}
\begin{eqnarray}
\sin 2 \tilde{\theta}^{}_{23}
\sin\tilde{\delta} = \sin 2\theta^{}_{23} \sin\delta \; ,
\label{eq:Toshev}
%     (3)
\end{eqnarray}
which directly links the fundamental flavor-mixing and CP-violating parameters
$\big(\theta^{}_{23}, \delta\big)$ to their effective counterparts
$\big(\tilde{\theta}^{}_{23}, \tilde{\delta}\big)$ in matter. Eq.~(\ref{eq:Toshev})
tells us that $\tilde{\theta}^{}_{23} = \pi/4$ and $\tilde{\delta} = \pm \pi/2$ are
a natural consequence of $\theta^{}_{23} = \pi/4$ and $\delta = \pm \pi/2$ in the
$\mu$-$\tau$ reflection symmetry limit~\cite{Harrison:2002et,Xing:2010ez,Xing:2015fdg}.

But the tree-level universality of weak neutral-current contributions to
${\cal V}^{}_\alpha$ in Eq.~(\ref{eq:H}) will be slightly broken once the one-loop
electroweak radiative corrections are taken into
account~\cite{Botella:1986wy,Mirizzi:2009td}. It is found that such quantum effects
are suppressed by the factors $G^{}_{\rm F} m^2_\alpha$ (for $\alpha = e, \mu, \tau$)
as compared with their corresponding tree-level terms and hence negligibly small in
both ${\cal V}^{}_e$ and ${\cal V}^{}_\mu$. In view of $m^{}_e \ll m^{}_\mu
\ll m^{}_\tau$~\cite{Zyla:2020zbs}, one obtains~\cite{Botella:1986wy,Mirizzi:2009td}
\begin{eqnarray}
{\cal V}^{}_e - {\cal V}^{}_\mu & = &
\sqrt{2} \hspace{0.04cm} G^{}_{\rm F} N^{}_e \; ,
\nonumber \\
{\cal V}^{}_\tau - {\cal V}^{}_\mu & = &
-\frac{3 G^2_{\rm F} m^2_\tau}{2 \pi^2} \left[\left(N^{}_p + N^{}_n \right)
\ln \frac{m^2_\tau}{m^2_W} + N^{}_p + \frac{2}{3} N^{}_n\right] \; \hspace{0.4cm}
\label{eq:one-loop}
%     (4)
\end{eqnarray}
in an excellent approximation, where $m^{}_\tau$ and $m^{}_W$ stand respectively
for the tau-lepton and $W$-boson masses. Taking $N^{}_p = N^{}_e$ for a neutral
medium and inputting $G^{}_{\rm F} \simeq 1.1664 \times 10^{-5}~{\rm GeV}^{-2}$,
$m^{}_W \simeq 80.377 ~{\rm GeV}$ and
$m^{}_\tau \simeq 1.777~{\rm GeV}$~\cite{Zyla:2020zbs}, we arrive at
\begin{eqnarray}
r \equiv \frac{{\cal V}^{}_\tau - {\cal V}^{}_\mu}
{{\cal V}^{}_e - {\cal V}^{}_\mu} & = &
\frac{3 G^{}_{\rm F} m^2_\tau}{2\sqrt{2} \hspace{0.05cm} \pi^2}
\left[\left(1 + \frac{N^{}_n}{N^{}_p}\right) \ln \frac{m^2_W}{m^2_\tau}
- 1 - \frac{2 N^{}_n}{3 N^{}_p} \right] \hspace{0.4cm}
\nonumber \\
& \simeq & \left\{ \begin{array}{l}
5.4 \times 10^{-5} \hspace{1cm} (N^{}_n/N^{}_p = 1) \; , \\
1.1 \times 10^{-4} \hspace{1cm} (N^{}_n/N^{}_p = 3) \; ,
\end{array} \right.
\label{eq:r}
%     (5)
\end{eqnarray}
where two typical numerical examples for the ratio $N^{}_n/N^{}_p$ have been given
for illustration. After the one-loop quantum corrections to the matter
potential term are included, the effective Hamiltonian in Eq.~(\ref{eq:H}) can be
rewritten as
\begin{eqnarray}
{\cal H}^{}_{\rm eff} = \frac{1}{2 E} \left[
U D \hspace{0.04cm} U^\dagger +
a \left(\begin{matrix} 1 & 0 & 0 \cr
0 & 0 & 0 \cr 0 & 0 & r \end{matrix}\right)\right] \equiv
\frac{1}{2 E} \hspace{0.07cm} V \widetilde{D} \hspace{0.04cm}
V^\dagger \; ,
\label{eq:H2}
%     (6)
\end{eqnarray}
where $D \equiv {\rm Diag}\big\{m^2_1, m^2_2, m^2_3\big\}$,
$\widetilde{D} \equiv {\rm Diag}\big\{\widetilde{m}^2_1, \widetilde{m}^2_2,
\widetilde{m}^2_3\big\}$, and $a = 2\sqrt{2} \hspace{0.04cm} G^{}_{\rm F} N^{}_e E$ is
referred to as the matter parameter.
It becomes obvious that the $O^{}_{23}$, $O^{}_\delta$
and $O^\dagger_\delta$ components of $U$ in Eq.~(\ref{eq:U}) do not commute any more
with the matter potential term in Eq.~(\ref{eq:H2}), and hence the elegant
tree-level Toshev relation in Eq.~(\ref{eq:Toshev}) should have no reason to hold at
the one-loop level.

Different from Ref.~\cite{Zhu:2020wuy}, where the properties of
$\big| V^{}_{\alpha i}\big|$ (for $\alpha = e, \mu, \tau$ and $i = 1, 2, 3$)
in the $r \neq 0$ but $a \to \infty$ limit are carefully discussed, the present
paper is intended to examine to what extent the tree-level Toshev relation is
modified by the one-loop radiative effects characterized by $r \neq 0$ for
finite $a$. Although such a correction is expected to
be negligible in most cases, it is likely to be significant when the neutrino beam
travels in dense matter. We find that this is indeed the case. Our approach is to
derive a set of differential equations of $\widetilde{\Delta}^{}_{ij}$,
$\tilde{\theta}^{}_{ij}$ and $\tilde{\delta}$ against the matter variable $a$ in
the $r \neq 0$ case, from which a new Toshev-like relation between the fundamental
quantities $\big(\theta^{}_{23}, \delta\big)$ and their matter-corrected counterparts
$\big(\tilde{\theta}^{}_{23}, \tilde{\delta}\big)$ is established. The running
behaviors of $\tilde{\theta}^{}_{ij}$ (for $ij = 12, 13, 23$) and $\tilde{\delta}$
with respect to $a$, together with an appreciable breaking effect of the tree-level
Toshev relation in dense matter, are numerically illustrated.

\section{Analytical calculations}

Eq.~(\ref{eq:H}) or (\ref{eq:H2})
shows that the effective Hamiltonian responsible for neutrino oscillations
in matter is exactly of the same form as that in vacuum, and hence the relevant effective
physical quantities evolving with the scale-like matter parameter $a$ can automatically
return to their fundamental counterparts in the $a \to 0$ limit. This observation reminds
us that the powerful renormalization-group-equation (RGE)
tool~\cite{StueckelbergdeBreidenbach:1952pwl,Gell-Mann:1954yli,Wilson:1971bg}
is actually applicable to describing the matter effects on neutrino masses and flavor
mixing parameters (see, e.g., Refs.~\cite{Chiu:2010da,Chiu:2017ckv,Xing:2018lob,
Wang:2019yfp,Wang:2019dal,Zhou:2020iei,Zeng:2022rxm}). Using such a RGE-like tool,
we differentiate both sides of Eq.~(\ref{eq:H2}) with respect to $a$ and arrive at
\begin{eqnarray}
\dot{\widetilde{D}} + \left[V^\dagger \dot{V},
\widetilde{D}\right] = V^\dagger
\left(\begin{matrix} 1 & 0 & 0 \cr 0 & 0 & 0 \cr 0 & 0 & r \end{matrix}\right)
V = \left(\begin{matrix}
\big|V^{}_{e 1}\big|^2 &
V^*_{e 1} V^{}_{e 2} &
V^*_{e 1} V^{}_{e 3} \cr
V^*_{e 2} V^{}_{e 1} &
\big|V^{}_{e 2}\big|^2 &
V^*_{e 2} V^{}_{e 3} \cr
V^*_{e 3} V^{}_{e 1} &
V^*_{e 3} V^{}_{e 2} &
\big|V^{}_{e 3}\big|^2
\end{matrix}\right) +
r \left(\begin{matrix}
\big|V^{}_{\tau 1}\big|^2 &
V^*_{\tau 1} V^{}_{\tau 2} &
V^*_{\tau 1} V^{}_{\tau 3} \cr
V^*_{\tau 2} V^{}_{\tau 1} &
\big|V^{}_{\tau 2}\big|^2 &
V^*_{\tau 2} V^{}_{\tau 3} \cr
V^*_{\tau 3} V^{}_{\tau 1} &
V^*_{\tau 3} V^{}_{\tau 2} &
\big|V^{}_{\tau 3}\big|^2
\end{matrix}\right) \; ,
\label{eq:RGE}
%     (7)
\end{eqnarray}
in which the overhead dot denotes the derivative of a matter-corrected quantity,
and the square brackets represents the commutator of two matrices
(i.e., $\big[A, B\big] \equiv AB - BA$). Then the diagonal and off-diagonal parts
of Eq.~(\ref{eq:RGE}) lead us to
\begin{eqnarray}
\dot{\widetilde{\Delta}}^{}_{ij} = \big|V^{}_{e i}\big|^2 -
\big|V^{}_{e j}\big|^2 + r \left(\big|V^{}_{\tau i}\big|^2 -
\big|V^{}_{\tau j}\big|^2\right) \; ,
\label{eq:RGE-mass-squared}
%     (8)
\end{eqnarray}
and
\begin{eqnarray}
\sum_{\alpha} V^*_{\alpha i} \dot{V}^{}_{\alpha j} =
\frac{V^*_{e i} V^{}_{e j} + r V^*_{\tau i}
V^{}_{\tau j}}{\widetilde{\Delta}^{}_{ji}}  \; ,
\label{eq:RGE2}
%     (9)
\end{eqnarray}
where $\widetilde{\Delta}^{}_{ij} \equiv \widetilde{m}^2_i - \widetilde{m}^2_j$
and $i \neq j$ (for $i, j = 1, 2, 3$). Thanks to the unitarity of $V$,
we simply have
\begin{eqnarray}
\sum_{i} \left(V^*_{\alpha i} \dot{V}^{}_{\beta i}
+ \dot{V}^*_{\alpha i} V^{}_{\beta i}\right)
= \sum_{\alpha} \left(V^*_{\alpha i} \dot{V}^{}_{\alpha j}
+ \dot{V}^*_{\alpha i} V^{}_{\alpha j}\right) = 0 \; ,
\label{eq:unitarity}
%     (10)
\end{eqnarray}
no matter whether $\alpha = \beta$ (or $i = j$) holds or not for the first
(or second) equalities. Now let us consider the orthogonality relation
\begin{eqnarray}
\sum_{j \neq i} V^*_{\alpha j} V^{}_{\beta j} = \delta^{}_{\alpha \beta}
- V^*_{\alpha i} V^{}_{\beta i} \; ,
\label{eq:orthogonality}
%     (11)
\end{eqnarray}
multiply both of its sides with $\dot{V}^{}_{\alpha i}$ and sum over the flavor index
$\alpha$. Then we obtain
\begin{eqnarray}
\dot{V}^{}_{\beta i} = \sum_\alpha \dot{V}^{}_{\alpha i} V^*_{\alpha i} V^{}_{\beta i}
+ \sum_{j\neq i} \frac{\left(V^{}_{ei} V^*_{e j} + r V^*_{\tau i} V^{}_{\tau j}\right)
V^{}_{\beta j}}{\widetilde{\Delta}^{}_{ij}} \; ,
\label{eq:V}
%     (12)
\end{eqnarray}
with the help of Eq.~(\ref{eq:RGE2}). This expression allows us to calculate the
differentials of nine squared moduli $X^{}_{\alpha i} \equiv \big|V^{}_{\alpha i}\big|^2$
against the matter parameter $a$ as follows:
\begin{eqnarray}
\dot{X}^{}_{\alpha i} = \dot{V}^*_{\alpha i} V^{}_{\alpha i}
+ V^*_{\alpha i} \dot{V}^{}_{\alpha i}
= 2 \sum_{j\neq i} \frac{{\rm Re}\big(V^{}_{e i} V^{}_{\alpha j}
V^*_{e j} V^*_{\alpha i}\big) + r {\rm Re}\big(V^{}_{\tau i} V^{}_{\alpha j}
V^*_{\tau j} V^*_{\alpha i}\big)}{\widetilde{\Delta}^{}_{ij}} \; .
\label{eq:V2}
%     (13)
\end{eqnarray}
If the one-loop contribution characterized by $r$ is switched off,
Eqs.~(\ref{eq:RGE-mass-squared})---(\ref{eq:V2}) will reproduce the previous
tree-level results obtained in Ref.~\cite{Xing:2018lob}.

Besides the rephasing invariants $X^{}_{\alpha i}$ in matter, the strength of
leptonic CP violation in neutrino oscillations is measured by the effective
Jarlskog invariant $\tilde{J}$ defined through~\cite{Jarlskog:1985ht}
\begin{eqnarray}
{\rm Im}\big(V^{}_{\alpha i} V^{}_{\beta j} V^*_{\alpha j} V^*_{\beta i}\big)
= \tilde{J} \sum_\gamma \sum_k \epsilon^{}_{\alpha\beta\gamma}
\epsilon^{}_{ijk} \; ,
\label{eq:J}
%     (14)
\end{eqnarray}
where $\epsilon^{}_{\alpha\beta\gamma}$ and $\epsilon^{}_{ijk}$ are the
three-dimensional Levi-Civita symbols with the Greek and Latin subscripts
running respectively over $(e, \mu, \tau)$ and $(1, 2, 3)$. The relationship
of $\tilde{J}$ with its fundamental counterpart $J$ in vacuum, which is
defined in the same manner as $\tilde{J}$ in Eq.~(\ref{eq:J}),
is known as the Naumov relation~\cite{Naumov:1991ju,Harrison:1999df,Xing:2001bg}
\begin{eqnarray}
\frac{\tilde{J}}{J} = \frac{\Delta^{}_{21} \Delta^{}_{31} \Delta^{}_{32}}
{\widetilde{\Delta}^{}_{21} \widetilde{\Delta}^{}_{31}
\widetilde{\Delta}^{}_{32}} \; ,
\label{eq:Naumov}
%     (15)
\end{eqnarray}
where $\Delta^{}_{ij} \equiv m^2_i - m^2_j$ (for $i, j = 1, 2, 3$). Our next
step is to show that this tree-level relation holds at the one-loop
level by starting from $\tilde{J} = {\rm Im}\big(V^{}_{\tau 1}
V^{}_{e 2} V^*_{\tau 2} V^*_{e 1}\big)$. With the help of Eq.~(\ref{eq:V}),
we find
\begin{eqnarray}
\dot{\tilde{J}} & = & {\rm Im} \big(\dot{V}^{}_{\tau 1}
V^{}_{e 2} V^*_{\tau 2} V^*_{e 1}\big) + {\rm Im} \big(V^{}_{\tau 1}
V^{}_{e 2} V^*_{\tau 2} \dot{V}^*_{e 1}\big)
+ {\rm Im} \big(V^{}_{\tau 1} \dot{V}^{}_{e 2} V^*_{\tau 2} V^*_{e 1}\big)
+ {\rm Im} \big(V^{}_{\tau 1} V^{}_{e 2} \dot{V}^*_{\tau 2} V^*_{e 1}\big)
\nonumber \\
& = & \tilde{J} \sum_{j > i} \frac{\big|V^{}_{e i}\big|^2 - \big|V^{}_{e j}\big|^2
+ r \left(\big|V^{}_{\tau i}\big|^2 - \big|V^{}_{\tau j}\big|^2\right)}
{\widetilde{\Delta}^{}_{ji}} \; .
\label{eq:J-equation}
%     (16)
\end{eqnarray}
Combining this equation with Eq.~(\ref{eq:RGE-mass-squared}), we simply arrive at
\begin{eqnarray}
\frac{{\rm d}}{{\rm d}a} \ln \left(\tilde{J} \widetilde{\Delta}^{}_{21}
\widetilde{\Delta}^{}_{31} \widetilde{\Delta}^{}_{32}\right) =
\frac{\dot{\tilde{J}}}{\tilde{J}} +
\frac{\dot{\widetilde{\Delta}}^{}_{21}}{\widetilde{\Delta}^{}_{21}} +
\frac{\dot{\widetilde{\Delta}}^{}_{31}}{\widetilde{\Delta}^{}_{31}} +
\frac{\dot{\widetilde{\Delta}}^{}_{32}}{\widetilde{\Delta}^{}_{32}} = 0 \; .
\label{eq:Naumov-proof}
%     (17)
\end{eqnarray}
This result indicates that the combination $\tilde{J} \widetilde{\Delta}^{}_{21}
\widetilde{\Delta}^{}_{31} \widetilde{\Delta}^{}_{32}$ is actually independent of the
matter parameter $a$, and hence it is just equal to $J \Delta^{}_{21} \Delta^{}_{31}
\Delta^{}_{32}$ at $a = 0$ (i.e., in vacuum); namely, Eq.~(\ref{eq:Naumov}) is valid.
We conclude that the intriguing Naumov relation in Eq.~(\ref{eq:Naumov}) {\it does}
hold at the one-loop level with $r \neq 0$, simply because in this case the matter
potential in Eq.~(\ref{eq:H2}) remains diagonal.

We proceed to derive the differential equations of $\tilde{\theta}^{}_{ij}$ (for
$ij = 12, 13, 23$) and $\tilde{\delta}$ with respect to $a$ in the standard
parametrization of $V$ which is exactly parallel to the one of $U$ in Eq.~(\ref{eq:U}).
Namely, $V$ can be expressed as $V = \widetilde{P}^{}_l \widetilde{O}^{}_{23}
\widetilde{O}^{}_\delta \widetilde{O}^{}_{13} \widetilde{O}^{\dagger}_\delta
\widetilde{O}^{}_{12} \widetilde{P}^{}_\nu$. Note that $\widetilde{P}^{}_\nu$ disappears
in the product $V \widetilde{D} V^\dagger$, and thus it does not affect the effective
Hamiltonian ${\cal H}^{}_{\rm eff}$ in Eq.~(\ref{eq:H}) or Eq.~(\ref{eq:H2}). Without
loss of generality, here we simply ignore $\widetilde{P}^{}_\nu$ and take
$\widetilde{P}^{}_l = {\rm Diag}\big\{e^{{\rm i}\phi^{}_e}, e^{{\rm i}\phi^{}_\mu}, 1\big\}$.
In this case we have
$V^\dagger \dot{V} = U^{\prime \dagger} \big(\widetilde{P}^{\dagger}_l
\dot{\widetilde{P}}^{}_l\big) U^\prime + U^{\prime \dagger} \dot{U}^\prime$ with
$U^\prime \equiv \widetilde{O}^{}_{23} \widetilde{O}^{}_\delta \widetilde{O}^{}_{13} \widetilde{O}^{\dagger}_\delta \widetilde{O}^{}_{12}$, from which we obtain
\begin{eqnarray}
\sum_\alpha U^{\prime *}_{\alpha i} \dot{U}^\prime_{\alpha j} + {\rm i}
\left[\dot{\phi}^{}_e U^{\prime *}_{e i} U^{\prime}_{e j} + \dot{\phi}^{}_\mu
U^{\prime *}_{\mu i} U^{\prime}_{\mu j} \right] =
\frac{U^{\prime *}_{e i} U^{\prime}_{ej} + r U^{\prime *}_{\tau i} U^{\prime}_{\tau j}}
{\widetilde{\Delta}^{}_{ji}} \; ,
\label{eq:parametrization+phases}
%     (18)
\end{eqnarray}
where $i \neq j$ is required. The diagonal part of this equation allows us to
reproduce Eq.~(\ref{eq:RGE-mass-squared}), and its off-diagonal part leads us to
the following differential equations for the three effective flavor mixing angles
and the effective CP-violating phase:
\begin{eqnarray}
\dot{\tilde{\theta}}^{}_{12} & = &
\tilde{c}^{}_{12} \tilde{s}^{}_{12} \left(\frac{\tilde{c}^2_{13}}{\widetilde{\Delta}^{}_{21}}
- \frac{\tilde{s}^2_{13} \widetilde{\Delta}^{}_{21}}{\widetilde{\Delta}^{}_{31}
\widetilde{\Delta}^{}_{32}}\right)
+ r \left[\frac{\left(\tilde{c}^2_{12}
- \tilde{s}^2_{12}\right) \tilde{s}^{}_{13} \tilde{c}^{}_{23} \tilde{s}^{}_{23}
\tilde{c}^{}_\delta + \tilde{c}^{}_{12} \tilde{s}^{}_{12} \left(\tilde{s}^2_{13} \tilde{c}^2_{23}
- \tilde{s}^2_{23}\right)}{\widetilde{\Delta}^{}_{21}} \right.
\nonumber \\
&& + \left. \frac{\tilde{s}^2_{12} \tilde{s}^{}_{13} \tilde{c}^{}_{23} \tilde{s}^{}_{23}
\tilde{c}^{}_\delta}{\widetilde{\Delta}^{}_{31}} +
\frac{\tilde{c}^2_{12} \tilde{s}^{}_{13} \tilde{c}^{}_{23} \tilde{s}^{}_{23}
\tilde{c}^{}_\delta}{\widetilde{\Delta}^{}_{32}} +
\frac{\tilde{c}^{}_{12} \tilde{s}^{}_{12} \tilde{s}^{2}_{13} \tilde{c}^{2}_{23}
\widetilde{\Delta}^{}_{21}}{\widetilde{\Delta}^{}_{31} \widetilde{\Delta}^{}_{32}}\right] \; ,
\nonumber \\
\dot{\tilde{\theta}}^{}_{13} & = &
\tilde{c}^{}_{13} \tilde{s}^{}_{13} \left(\frac{\tilde{c}^2_{12}}{\widetilde{\Delta}^{}_{31}}
+ \frac{\tilde{s}^2_{12}}{\widetilde{\Delta}^{}_{32}}\right)
- r \tilde{c}^{}_{13} \tilde{c}^{}_{23} \left[\frac{\tilde{s}^{}_{13} \tilde{c}^{}_{23}}
{\widetilde{\Delta}^{}_{32}} + \frac{\tilde{c}^{}_{12}
\left(\tilde{s}^{}_{12} \tilde{s}^{}_{23} \tilde{c}^{}_\delta -
\tilde{c}^{}_{12} \tilde{s}^{}_{13} \tilde{c}^{}_{23}\right)
\widetilde{\Delta}^{}_{21}}{\widetilde{\Delta}^{}_{31} \widetilde{\Delta}^{}_{32}}\right] \; ,
\hspace{0.4cm}
\nonumber \\
\dot{\tilde{\theta}}^{}_{23} & = &
\frac{\tilde{c}^{}_{12} \tilde{s}^{}_{12} \tilde{s}^{}_{13} \tilde{c}^{}_\delta \widetilde{\Delta}^{}_{21}}{\widetilde{\Delta}^{}_{31} \widetilde{\Delta}^{}_{32}}
- r \tilde{c}^{}_{23} \left[\frac{\tilde{s}^{2}_{12} \tilde{s}^{}_{23}}
{\widetilde{\Delta}^{}_{31}} + \frac{\tilde{c}^{2}_{12} \tilde{s}^{}_{23}}
{\widetilde{\Delta}^{}_{32}} + \frac{\tilde{c}^{}_{12} \tilde{s}^{}_{12} \tilde{s}^{}_{13}
\tilde{c}^{}_{23} \tilde{c}^{}_\delta \widetilde{\Delta}^{}_{21}}
{\widetilde{\Delta}^{}_{31} \widetilde{\Delta}^{}_{32}}\right] \; ,
\label{eq:angles}
%     (19)
\end{eqnarray}
and
\begin{eqnarray}
\dot{\tilde{\delta}} & = & - \hspace{0.1cm}
\frac{\tilde{c}^{}_{12} \tilde{s}^{}_{12} \tilde{s}^{}_{13} \left(\tilde{c}^2_{23}
- \tilde{s}^2_{23}\right) \tilde{s}^{}_\delta \widetilde{\Delta}^{}_{21}}
{\tilde{c}^{}_{23} \tilde{s}^{}_{23} \widetilde{\Delta}^{}_{31} \widetilde{\Delta}^{}_{32}}
- r \tilde{c}^{}_{23} \tilde{s}^{}_\delta
\left[\frac{\tilde{s}^{}_{13} \tilde{s}^{}_{23}}
{\tilde{c}^{}_{12} \tilde{s}^{}_{12} \widetilde{\Delta}^{}_{21}} -
\frac{\tilde{s}^{3}_{12} \tilde{s}^{}_{13} \tilde{s}^{}_{23}}
{\tilde{c}^{}_{12} \widetilde{\Delta}^{}_{31}}
+ \frac{\tilde{c}^{3}_{12} \tilde{s}^{}_{13} \tilde{s}^{}_{23}}
{\tilde{s}^{}_{12} \widetilde{\Delta}^{}_{32}} \right.
\hspace{0.4cm}
\nonumber \\
&& - \left. \frac{\tilde{c}^{}_{12} \tilde{s}^{}_{12}
\left(\tilde{s}^{2}_{13} \tilde{c}^2_{23} + \tilde{c}^2_{13} \tilde{s}^2_{23}
\right) \widetilde{\Delta}^{}_{21}}{\tilde{s}^{}_{13} \tilde{s}^{}_{23}
\widetilde{\Delta}^{}_{31} \widetilde{\Delta}^{}_{32}}\right] \; ,
\label{eq:phase}
%     (20)
\end{eqnarray}
where $\tilde{c}^{}_{ij} \equiv \cos\tilde{\theta}^{}_{ij}$,
$\tilde{s}^{}_{ij} \equiv \sin\tilde{\theta}^{}_{ij}$,
$\tilde{c}^{}_\delta \equiv \cos\tilde{\delta}$ and
$\tilde{s}^{}_\delta \equiv \sin\tilde{\delta}$ have been defined. In comparison, the
two unphysical phases $\phi^{}_e$ and $\phi^{}_\mu$ evolve with $a$ as follows:
\begin{eqnarray}
\dot{\phi}^{}_e & = &
-\frac{\tilde{c}^{}_{12} \tilde{s}^{}_{12} \tilde{s}^{}_{13} \tilde{c}^{}_{23}
\tilde{s}^{}_\delta \widetilde{\Delta}^{}_{21}}
{\tilde{s}^{}_{23} \widetilde{\Delta}^{}_{31} \widetilde{\Delta}^{}_{32}}
- r \tilde{s}^{}_{13} \tilde{c}^{}_{23} \tilde{s}^{}_\delta
\left[\frac{\tilde{s}^{}_{23}}{\tilde{c}^{}_{12} \tilde{s}^{}_{12}
\widetilde{\Delta}^{}_{21}} - \frac{\tilde{s}^{3}_{12} \tilde{s}^{}_{23}}
{\tilde{c}^{}_{12} \widetilde{\Delta}^{}_{31}}
+ \frac{\tilde{c}^{3}_{12} \tilde{s}^{}_{23}}
{\tilde{s}^{}_{12} \widetilde{\Delta}^{}_{32}}
- \frac{\tilde{c}^{}_{12} \tilde{s}^{}_{12} \tilde{c}^2_{23}
\widetilde{\Delta}^{}_{21}} {\tilde{s}^{}_{23}
\widetilde{\Delta}^{}_{31} \widetilde{\Delta}^{}_{32}}\right] \; ,
\nonumber \\
\dot{\phi}^{}_\mu & = &
-\frac{\tilde{c}^{}_{12} \tilde{s}^{}_{12} \tilde{s}^{}_{13} \tilde{s}^{}_\delta
\widetilde{\Delta}^{}_{21}}{\tilde{c}^{}_{23} \tilde{s}^{}_{23}
\widetilde{\Delta}^{}_{31} \widetilde{\Delta}^{}_{32}}
+ r \left[\frac{\tilde{c}^{}_{12} \tilde{s}^{}_{12} \tilde{s}^{}_{13}
\tilde{c}^{}_{23} \tilde{s}^{}_\delta \widetilde{\Delta}^{}_{21}}{\tilde{s}^{}_{23}
\widetilde{\Delta}^{}_{31} \widetilde{\Delta}^{}_{32}}\right] \; .
\label{eq:unphysical-phase}
%     (21)
\end{eqnarray}
Although $\phi^{}_e$ and $\phi^{}_\mu$ have no physical significance, they cannot be
ignored in deriving the differential equations of those physical parameters in
Eqs.~(\ref{eq:angles}) and (\ref{eq:phase}) so as to keep the self-consistency
of our calculations. Two immediate comments on the evolution of $\tilde{\delta}$
and $\tilde{\theta}^{}_{23}$ with $a$ are in order.
\begin{itemize}
\item     Eq.~(\ref{eq:phase}) tells us that $\dot{\tilde{\delta}} \propto
\sin \tilde{\delta}$ holds at the one-loop level. This interesting proportionality is
fully consistent with $\tilde{J} = \tilde{c}^{}_{12} \tilde{s}^{}_{12} \tilde{c}^2_{13}
\tilde{s}^{}_{13} \tilde{c}^{}_{23} \tilde{s}^{}_{23} \tilde{s}^{}_\delta
\propto J = c^{}_{12} s^{}_{12} c^2_{13} s^{}_{13} c^{}_{23} s^{}_{23} s^{}_\delta$
revealed by the one-loop Naumov relation as shown in Eq.~(\ref{eq:Naumov-proof}),
and it implies that a nontrivial effective CP-violating phase cannot be generated from
the matter-induced correction to $\delta = 0$ (or $\pi$).

\item     $\dot{\tilde{\theta}}^{}_{23} \propto \cos\tilde{\delta}$ and
$\dot{\tilde{\delta}} \propto \cos 2\tilde{\theta}^{}_{23}$ hold in the
$r = 0$ case, so $\theta^{}_{23} = \pi/4$ and $\delta = \pm\pi/2$ in vacuum
(i.e., $a = 0$) will automatically assure $\tilde{\theta}^{}_{23} = \pi/4$ and
$\tilde{\delta} = \pm \pi/2$ to hold in matter. In other words, the $\mu$-$\tau$
reflection symmetry of $U$ is respected by the matter effect at the tree level,
but it will be slightly broken at the one-loop level (i.e., $r \neq 0$). So the
Toshev relation is not expected to exactly hold at the one-loop level either.
\end{itemize}
The explicit running behaviors of $\tilde{\delta}$ and $\tilde{\theta}^{}_{ij}$
(for $ij = 12, 13, 23$) with respect to the matter parameter $a$ will be numerically
illustrated in the next section.

Using the last two equations obtained in Eq.~(\ref{eq:phase}), we may easily examine
to what extant the Toshev relation is broken at the one-loop level. We find
\begin{eqnarray}
\frac{{\rm d}}{{\rm d} a} \ln\left( \sin 2\tilde{\theta}^{}_{23} \sin \tilde{\delta}
\right) = \left(2\cot 2\tilde{\theta}^{}_{23}\right) \dot{\tilde{\theta}}^{}_{23} +
\left(\cot\tilde{\delta}\right) \dot{\tilde{\delta}}
= - r \kappa \; ,
\label{eq:dToshev}
%    (22)
\end{eqnarray}
where
\begin{eqnarray}
\kappa & = & \frac{\tilde{s}^{}_{13} \tilde{c}^{}_{23} \tilde{s}^{}_{23}
\tilde{c}^{}_\delta}{\tilde{c}^{}_{12} \tilde{s}^{}_{12} \widetilde{\Delta}^{}_{21}}
+ \frac{\tilde{c}^{}_{12} \tilde{s}^2_{12} \tilde{s}^{}_{13}
\left(\tilde{c}^2_{23} - \tilde{s}^2_{23}\right) + \tilde{s}^{}_{12}
\tilde{c}^{}_{23} \tilde{s}^{}_{23} \tilde{c}^{}_\delta \left(\tilde{c}^2_{12}
- \tilde{s}^2_{12} \tilde{s}^2_{13}\right)}{\tilde{c}^{}_{12} \tilde{s}^{}_{13}
\widetilde{\Delta}^{}_{31}}
\nonumber \\
&&
+ \left. \frac{\tilde{c}^{2}_{12} \tilde{s}^{}_{12} \tilde{s}^{}_{13}
\left(\tilde{c}^2_{23} - \tilde{s}^2_{23}\right) - \tilde{c}^{}_{12}
\tilde{c}^{}_{23} \tilde{s}^{}_{23} \tilde{c}^{}_\delta \left(\tilde{s}^2_{12}
- \tilde{c}^2_{12} \tilde{s}^2_{13}\right)}{\tilde{s}^{}_{12} \tilde{s}^{}_{13}
\widetilde{\Delta}^{}_{32}} \right. \;
\label{eq:kappa}
%    (23)
\end{eqnarray}
is a complicated nonlinear function of $a$. Although it is almost impossible to solve
Eq.~(\ref{eq:dToshev}) in an exact analytical way, a formal solution to this
equation can be expressed as
\begin{eqnarray}
R \equiv \frac{\sin 2\tilde{\theta}^{}_{23} \sin\tilde{\delta}}
{\sin 2\theta^{}_{23} \sin\delta} = \exp\left[-r \int^a_0 \kappa {\rm d} a\right] \; .
\label{eq:Toshev-one-loop}
%    (24)
\end{eqnarray}
It is obvious that $R = 1$ exactly holds either in the $a = 0$ case (i.e., in vacuum)
or in the $r = 0$ case (i.e., at the tree level of coherent forward neutrino
scattering with matter). Note that $a$ is proportional to $N^{}_e$ but $r$ is closely
associated with $N^{}_n/N^{}_p$, so a significant deviation of $R$ from one is still
possible for a small value of $r$ provided $a$ is large enough
%%%%%%%%%%%%%%%%%%%%%%%%%%%%%%%%%%%%%%%%%%%%%%%%%%%%%%%%%%%%%%%%%%%%%%%%%%%%%%%%%%%%%
\footnote{If $a$ is sufficiently large, however, there will be no good reason to neglect
the electroweak radiative correction to the tree-level matter potential of
${\cal H}^{}_{\rm eff}$ in Eq.~(\ref{eq:H2}) because it is the term $a r$ that
takes effect. This observation implies that some previous discussions about the
``asymptotic" behaviors of $\big|V^{}_{\alpha i}\big|$ (for $\alpha = e, \mu, \tau$
and $i = 1, 2, 3$) or $\big(\tilde{\theta}^{}_{12},
\tilde{\theta}^{}_{13}, \tilde{\theta}^{}_{23}, \tilde{\delta}\big)$ in the
$a \to \infty$ limit but {\it at the tree level}~(see, e.g.,
Refs.~\cite{Xing:2019owb,Luo:2019efb}) are likely problematic and even wrong.
The present work can therefore provide an important clarification in this regard.}.
%%%%%%%%%%%%%%%%%%%%%%%%%%%%%%%%%%%%%%%%%%%%%%%%%%%%%%%%%%%%%%%%%%%%%%%%%%%%%%%%%%%%%

So far we have only considered the case of a neutrino beam traveling in matter. When
an antineutrino beam is taken into account, the PMNS matrix $U$ and the matter
parameter $a$ in Eq.~(\ref{eq:H2}) should be replaced respectively with $U^*$ and $-a$
such that $a$ itself is universal for both neutrinos and antineutrinos. In this case
it is $V^*$ that effectively describes antineutrino oscillations in a medium. As a result,
the right-hand sides of Eqs.~(\ref{eq:RGE-mass-squared}) and (\ref{eq:angles}) should be
multiplied by a negative sign but Eq.~(\ref{eq:phase}) keeps unchanged when they are
applied to calculating the effective antineutrino oscillation parameters in matter.

\section{Numerical illustration}

To illustrate how the four effective flavor mixing parameters $\big(\tilde{\theta}^{}_{12},
\tilde{\theta}^{}_{13}, \tilde{\theta}^{}_{23}, \tilde{\delta}\big)$ and the three
effective neutrino mass-squared differences $\big(\widetilde{\Delta}^{}_{21},
\widetilde{\Delta}^{}_{31}, \widetilde{\Delta}^{}_{32}\big)$ evolve with
the matter parameter $a$ at the one-loop level, we typically choose
$r \simeq 5.4\times 10^{-5}$ for a neutral medium with $N^{}_n = N^{}_p = N^{}_e$ and
$r \simeq 1.1\times 10^{-4}$ for a more dense medium with $N^{}_n = 3N^{}_p = 3N^{}_e$
as estimated in Eq.~(\ref{eq:r}). The RGE-like differential equations obtained in
Eqs.~(\ref{eq:RGE-mass-squared}), (\ref{eq:angles}) and (\ref{eq:phase}) will be used
for doing the numerical calculations. We input the following best-fit values of the six
independent neutrino oscillation parameters extracted from current experimental data in vacuum~\cite{Gonzalez-Garcia:2021dve}: (1) as for the normal mass ordering (NMO) of three
active neutrinos (i.e., $m^{}_1 < m^{}_2 < m^{}_3$),
$\theta^{}_{12} \simeq 33.45^{\circ}$,
$\theta^{}_{13} \simeq 8.62^{\circ}$,
$\theta^{}_{23} \simeq 42.1^{\circ}$,
$\delta \simeq 230^{\circ}$,
$\Delta^{}_{21} \simeq 7.42 \times 10^{-5}~{\rm eV}^2$ and
$\Delta^{}_{31} \simeq 2.510 \times 10^{-3}~{\rm eV}^2$;
(2) as for the inverted mass ordering (IMO) of three active neutrinos (i.e.,
$m^{}_3 < m^{}_1 < m^{}_2$),
$\theta^{}_{12} \simeq 33.45^{\circ}$,
$\theta^{}_{13} \simeq 8.61^{\circ}$,
$\theta^{}_{23} \simeq 49.0^{\circ}$,
$\delta \simeq 278^{\circ}$,
$\Delta^{}_{21} \simeq 7.42\times 10^{-5}~{\rm eV}^2$ and
$\Delta^{}_{32} = -2.490 \times 10^{-3}~{\rm eV}^2$.
Our results are illustrated by
Figs.~\ref{fig:NMO-neutrino-angles}---\ref{fig:IMO-antineutrino-m2}, in which both
the NMO and IMO cases for both neutrino and antineutrino oscillations have been
taken into account
%%%%%%%%%%%%%%%%%%%%%%%%%%%%%%%%%%%%%%%%%%%%%%%%%%%%%%%%%%%%%%%%%%%%%%%%%%%%%%%%%%%%%%%%%%
\footnote{We have plotted these figures by allowing the matter parameter $a$ to change
from $10^{-7}~{\rm eV}^2$ to $10^{4}~{\rm eV}^2$, such that the behaviors of relevant
neutrino oscillation parameters evolving with $a$ at the one-loop level can be fully
exhibited. It is therefore reasonable to assume that these two ``endpoints" of
$a$ are numerically equivalent to the vacuum limit ($a \to 0$) and the dense matter limit
($a \to \infty$) to reveal the ``asymptotic" values of $\tilde{\theta}^{}_{12}$,
$\tilde{\theta}^{}_{13}$, $\tilde{\theta}^{}_{23}$ and $\tilde{\delta}$.}.
%%%%%%%%%%%%%%%%%%%%%%%%%%%%%%%%%%%%%%%%%%%%%%%%%%%%%%%%%%%%%%%%%%%%%%%%%%%%%%%%%%%%%%%%%%
Some brief discussions are in order.
%%%%%%%%%%%%%%%%%%%%%%%%%%%%%%%%% Figure 1 %%%%%%%%%%%%%%%%%%%%%%%%%%%%%%%%%%%%%%%%%%%%%%%
\begin{figure}[t]
\centering
\includegraphics[width=0.9\textwidth]{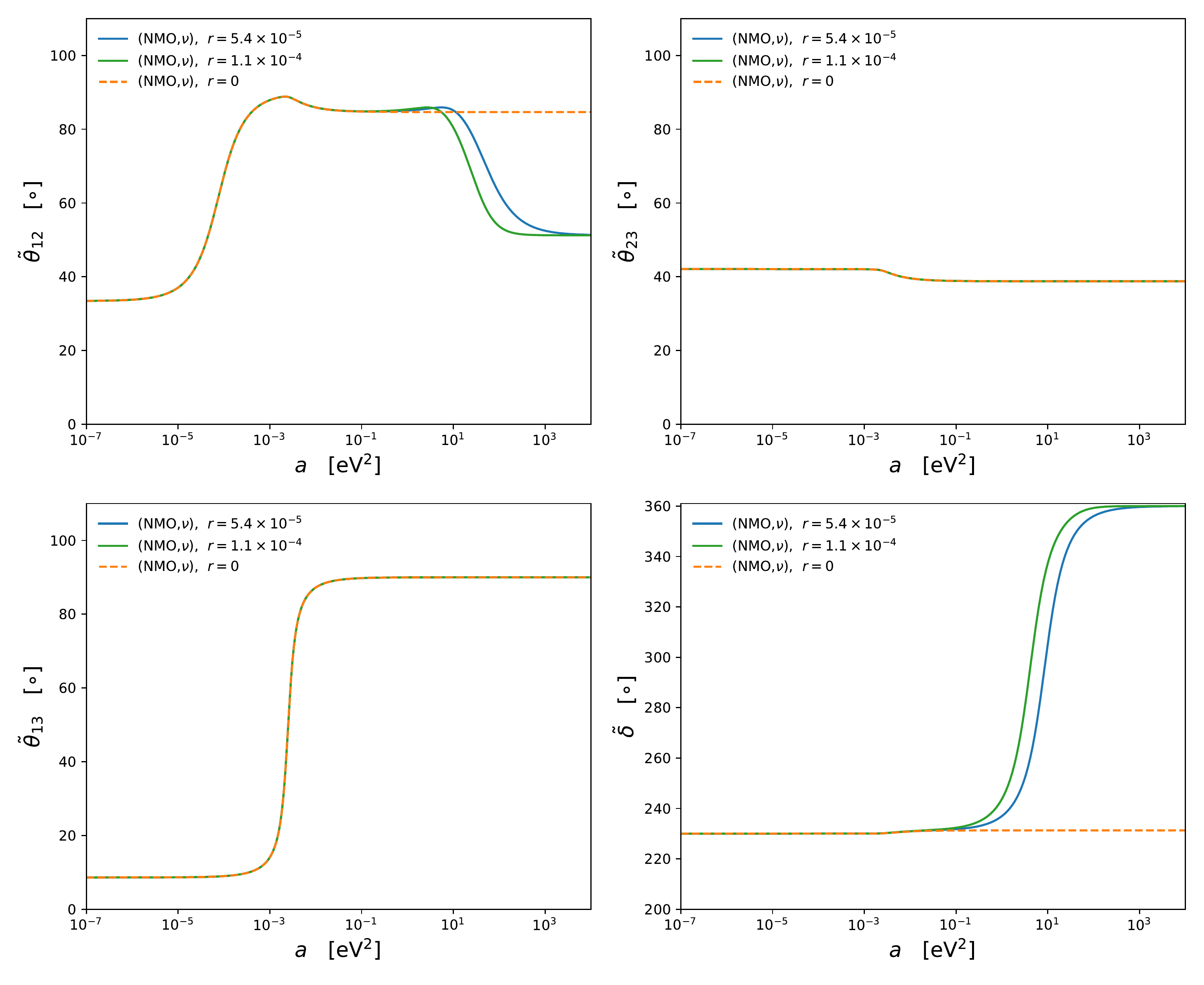}
\vspace{-0.6cm}
\caption{An illustration of the four effective flavor-mixing and CP-violating parameters
evolving with the matter parameter $a$ in the NMO case for neutrino oscillations.}
\label{fig:NMO-neutrino-angles}
\end{figure}
%%%%%%%%%%%%%%%%%%%%%%%%%%%%%%%%%%%%%%%%%%%%%%%%%%%%%%%%%%%%%%%%%%%%%%%%%%%%%%%%%%%%%%%%%%
%%%%%%%%%%%%%%%%%%%%%%%%%%%%%%%% Figure 2 %%%%%%%%%%%%%%%%%%%%%%%%%%%%%%%%%%%%%%%%%%%%%%%%
\begin{figure}[!h]
\centering
\includegraphics[width=1\textwidth]{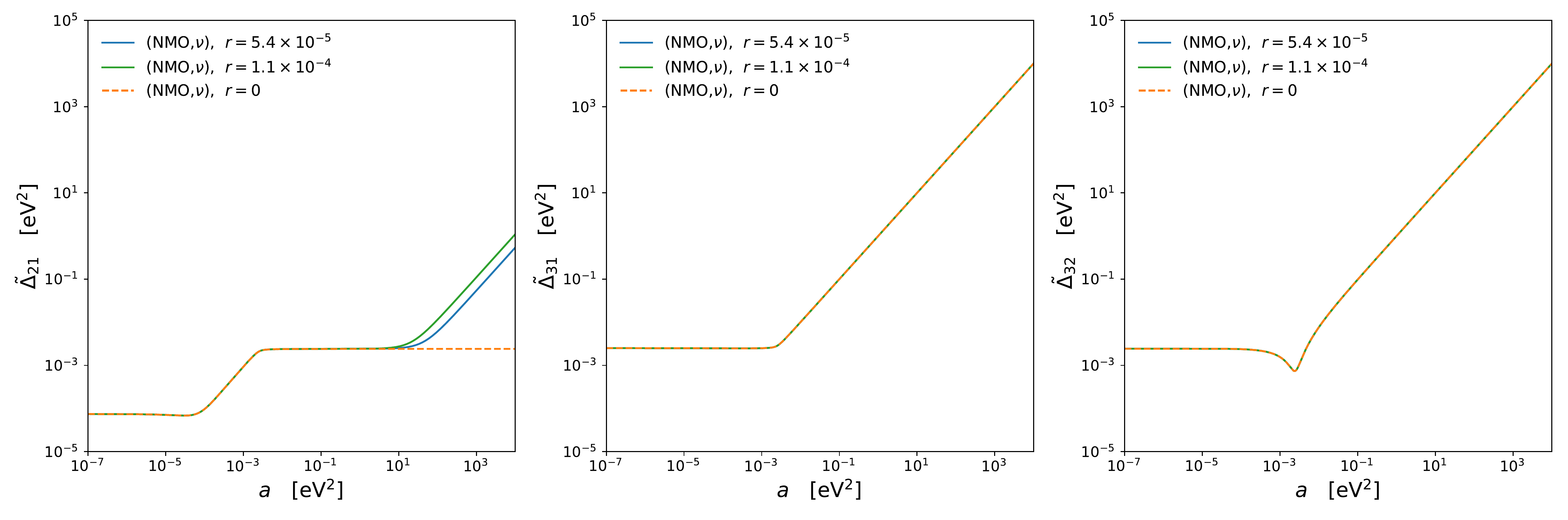}
\vspace{-1.1cm}
\caption{An illustration of the three effective neutrino mass-squared differences
evolving with the matter parameter $a$ in the NMO case for neutrino oscillations.}
\label{fig:NMO-neutrino-m2}
\end{figure}
%%%%%%%%%%%%%%%%%%%%%%%%%%%%%%%%%%%%%%%%%%%%%%%%%%%%%%%%%%%%%%%%%%%%%%%%%%%%%%%%%%%%%%%%%%

(1) {\it The NMO case for neutrino oscillations} (Figs.~\ref{fig:NMO-neutrino-angles}
and \ref{fig:NMO-neutrino-m2}). Switching off the one-loop radiative correction
to the matter potential of ${\cal H}^{}_{\rm eff}$ (i.e., taking $r = 0$), we have
confirmed that our numerical results for $\tilde{\theta}^{}_{ij}$, $\tilde{\delta}$ and
$\widetilde{\Delta}^{}_{ji}$ (for $ij = 12, 13, 23$) are fully compatible with those
obtained in Ref.~\cite{Xing:2018lob}, where all the salient features of these effective
neutrino oscillation parameters evolving with $a$ have been understood and interpreted
with the help of their corresponding differential equations. Here we do not repeat the
same discussions but focus on the new effect caused by $r \neq 0$. One can see that the
quantum effect characterized by $r \gtrsim 5.4 \times 10^{-5}$ becomes important for
$\tilde{\theta}^{}_{12}$ and $\tilde{\delta}$ when $a \gtrsim 0.1~{\rm eV}^2$ holds
in a dense medium. The reason is that both $\dot{\tilde{\theta}}^{}_{12}$ and
$\dot{\tilde{\delta}}$ contain the terms proportional to $r/\widetilde{\Delta}^{}_{21}$,
but $\widetilde{\Delta}^{}_{21}$ itself is not significantly enhanced by the matter
effect until $a \gtrsim 10~{\rm eV}^2$. Such a {\it delayed} matter-induced enhancement of
$\widetilde{\Delta}^{}_{21}$ implies that $\tilde{\theta}^{}_{12}$ and $\tilde{\delta}$
will finally approach their corresponding ``fixed points" when $a$ is much larger, as
can be seen in Fig.~\ref{fig:NMO-neutrino-angles}. In comparison, $\tilde{\theta}^{}_{13}$
and $\tilde{\theta}^{}_{23}$ are essentially insensitive to $r \neq 0$ because they
depend only upon $r/\widetilde{\Delta}^{}_{31}$ and $r/\widetilde{\Delta}^{}_{32}$ but
$\widetilde{\Delta}^{}_{31}$ and $\widetilde{\Delta}^{}_{32}$ themselves increase
rapidly when $a \gtrsim 10^{-2}~{\rm eV}^2$ holds.

Now let us explain why $\tilde{\theta}^{}_{12} \simeq 51.1^\circ$ holds as the one-loop
fixed point of $\tilde{\theta}^{}_{12}$ in the $a \to \infty$ limit. One the one hand,
we see $\tilde{\theta}^{}_{13} \to 90^\circ$ and $\tilde{\delta} \to 360^\circ$
(or equivalently, $\tilde{\delta} \to 0^\circ$) in this limit, and thus the dominant
term of $\dot{\tilde{\theta}}^{}_{12}$ in Eq.~(\ref{eq:angles}) can be simplified to
\begin{eqnarray}
\dot{\tilde{\theta}}^{}_{12} \simeq
\frac{r \sin 2\big(\tilde{\theta}^{}_{12} + \tilde{\theta}^{}_{23}\big)}
{2 \widetilde{\Delta}^{}_{21}} \to 0 \;
\label{eq:theta12}
%    (25)
\end{eqnarray}
when $\tilde{\theta}^{}_{12}$ approaches its fix point,
implying $\tilde{\theta}^{}_{12} + \tilde{\theta}^{}_{23} \simeq 90^\circ$ in this
special case. On the other hand, the asymptotic value of $\tilde{\theta}^{}_{23}$
can be determined from the tree-level formula
\begin{eqnarray}
\tilde{t}^{}_{23} \simeq \left|t^{}_{23} + e^{{\rm i}\delta}
\frac{\Delta^{}_{21}}{\Delta^{}_{31}} \cdot \frac{c^{}_{12} s^{}_{12}}
{s^{}_{13} c^2_{23}}\right| \;
\label{eq:theta23}
%    (26)
\end{eqnarray}
with $t^{}_{23} \equiv \tan\theta^{}_{23}$ and
$\tilde{t}^{}_{23} \equiv \tan\tilde{\theta}^{}_{23}$
in the approximation of $\Delta^{}_{21} \ll \Delta^{}_{31} \ll a$~\cite{Freund:2001pn,
Cervera:2000kp}, because the one-loop radiative correction to $\tilde{\theta}^{}_{23}$
is negligibly small. We obtain $\tilde{\theta}^{}_{23} \simeq 38.9^\circ$ from
Eq.~(\ref{eq:theta23}), and thus arrive at $\tilde{\theta}^{}_{12} \simeq 90^\circ
- \tilde{\theta}^{}_{23} \simeq 51.1^\circ$ as $a \to \infty$. This one-loop result is
remarkably different from the tree-level one obtained in Ref.~\cite{Xing:2018lob},
telling us why the one-loop radiative corrections to the matter potential of
${\cal H}^{}_{\rm eff}$ must be taken into account for neutrino
oscillations in dense matter.
%%%%%%%%%%%%%%%%%%%%%%%%%%%%%%%%%% Figure 3 %%%%%%%%%%%%%%%%%%%%%%%%%%%%%%%%%%%%%%%%%%%%
\begin{figure}[t]
\centering
\includegraphics[width=0.9\textwidth]{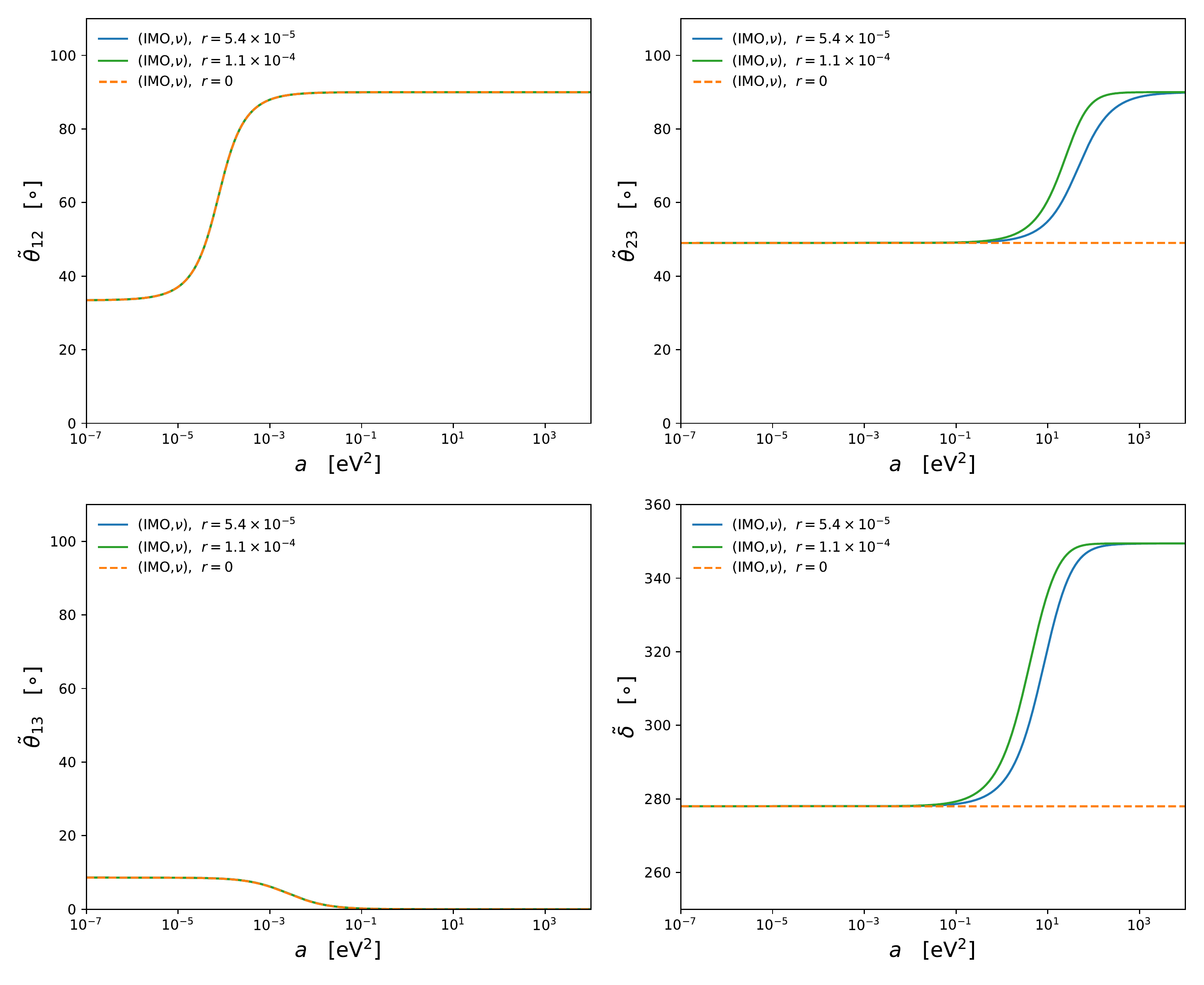}
\vspace{-0.6cm}
\caption{An illustration of the four effective flavor-mixing and CP-violating parameters
evolving with the matter parameter $a$ in the IMO case for neutrino oscillations.}
\label{fig:IMO-neutrino-angles}
\end{figure}
%%%%%%%%%%%%%%%%%%%%%%%%%%%%%%%%%%%%%%%%%%%%%%%%%%%%%%%%%%%%%%%%%%%%%%%%%%%%%%%%%%%%%%%%
%%%%%%%%%%%%%%%%%%%%%%%%%%%%%%%%% Figure 4 %%%%%%%%%%%%%%%%%%%%%%%%%%%%%%%%%%%%%%%%%%%%%
\begin{figure}[!h]
\centering
\includegraphics[width=1\textwidth]{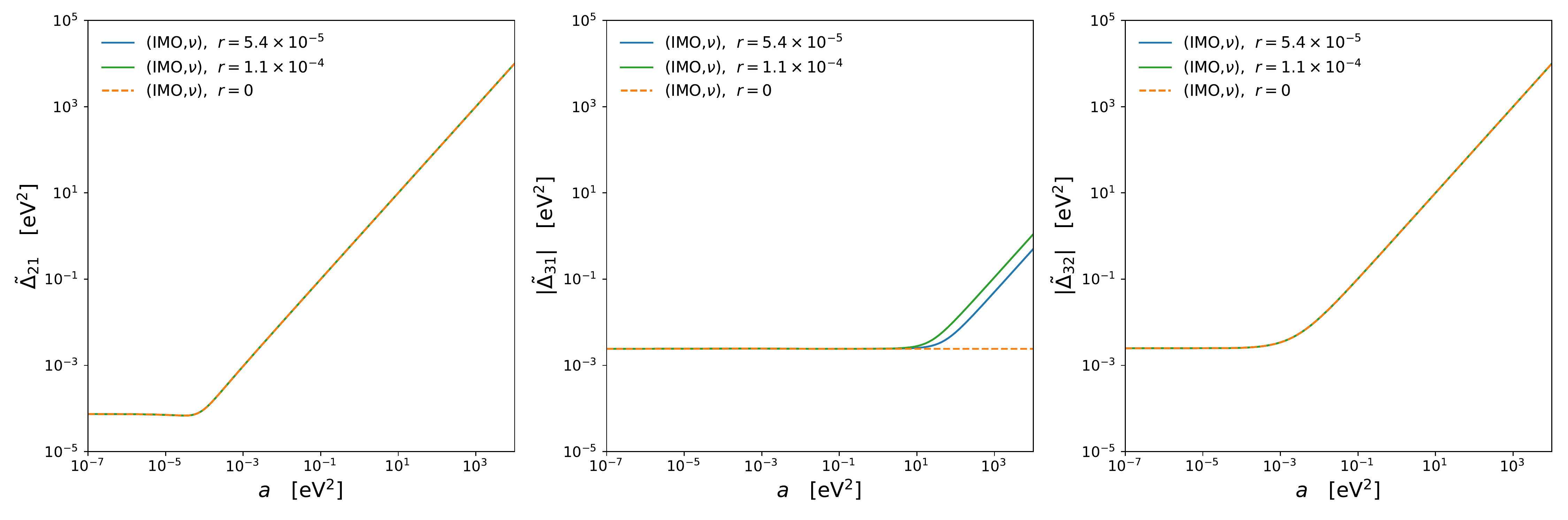}
\vspace{-1.1cm}
\caption{An illustration of the three effective neutrino mass-squared differences
evolving with the matter parameter $a$ in the IMO case for neutrino oscillations.}
\label{fig:IMO-neutrino-m2}
\end{figure}
%%%%%%%%%%%%%%%%%%%%%%%%%%%%%%%%%%%%%%%%%%%%%%%%%%%%%%%%%%%%%%%%%%%%%%%%%%%%%%%%%%%%%%%%

(2) {\it The IMO case for neutrino oscillations} (Figs.~\ref{fig:IMO-neutrino-angles}
and \ref{fig:IMO-neutrino-m2}). The situation is quite different in this case.
Fig.~\ref{fig:IMO-neutrino-angles} shows that $\tilde{\theta}^{}_{12}$ and
$\tilde{\theta}^{}_{13}$ are insensitive to the one-loop correction characterized
by $r \neq 0$ and have reached their corresponding asymptotic values $90^\circ$ and
$0^\circ$ as $a \simeq 0.1 ~{\rm eV}^2$~\cite{Xing:2018lob}, but
$\tilde{\theta}^{}_{23}$ and $\tilde{\delta}$ are very sensitive to this quantum effect
when $a\gtrsim 0.1 ~{\rm eV}^2$.
Note that $\big|\widetilde{\Delta}_{32}\big| \simeq \widetilde{\Delta}_{21}\gg \big|
\widetilde{\Delta}_{31}\big|$ holds for $a\gtrsim 0.1~{\rm eV}^2$, as
shown in Fig.~\ref{fig:IMO-neutrino-m2}, and thus the evolution of
$\tilde{\theta}^{}_{23}$ and $\tilde{\delta}$ with $a\gtrsim 0.1~{\rm eV}^2$ is
dominated by the terms proportional to $r/\widetilde{\Delta}^{}_{31}$ in
Eq.~(\ref{eq:angles}). To be explicit,
\begin{eqnarray}
\dot{\tilde{\theta}}^{}_{23} \simeq - \frac{r \sin 2\tilde{\theta}^{}_{23}}
{2 \widetilde{\Delta}^{}_{31}} \to 0 \;
\label{eq:theta23IMO}
%    (27)
\end{eqnarray}
in this region, leading to the fixed point $\tilde{\theta}^{}_{23} \simeq 90^\circ$.
In comparison, it is not straightforward to analytically understand the running behavior
of $\tilde{\delta}$ nearby its asymptotic value $\tilde{\delta}\simeq 349.4^{\circ}$, as
$\tilde{c}^{}_{12} \to 0$ and $\tilde{s}^{}_{13} \to 0$ appear in the denominators
of three terms of $\dot{\tilde{\delta}}$ given by Eq.~(\ref{eq:angles}) although
their divergent effects are not only compensated by $\tilde{c}^{}_{23} \to 0$ but
also suppressed respectively by $\tilde{s}^{}_{13} \to 0$  and $\tilde{c}^{}_{12} \to 0$
in the corresponding numerators. Of course, $\tilde{J} \to 0$ holds even in the limit of
$\tilde{\delta} \to 349.4^{\circ}$ because its vanishing is simultaneously governed by
$\tilde{c}^{}_{12} \to 0$, $\tilde{s}^{}_{13} \to 0$ and $\tilde{c}^{}_{23} \to 0$.
%%%%%%%%%%%%%%%%%%%%%%%%%%%%%%%%%%%% Figure 5 %%%%%%%%%%%%%%%%%%%%%%%%%%%%%%%%%%%%%%%%%%%%%%%
\begin{figure}[t]
\centering
\includegraphics[width=0.9\textwidth]{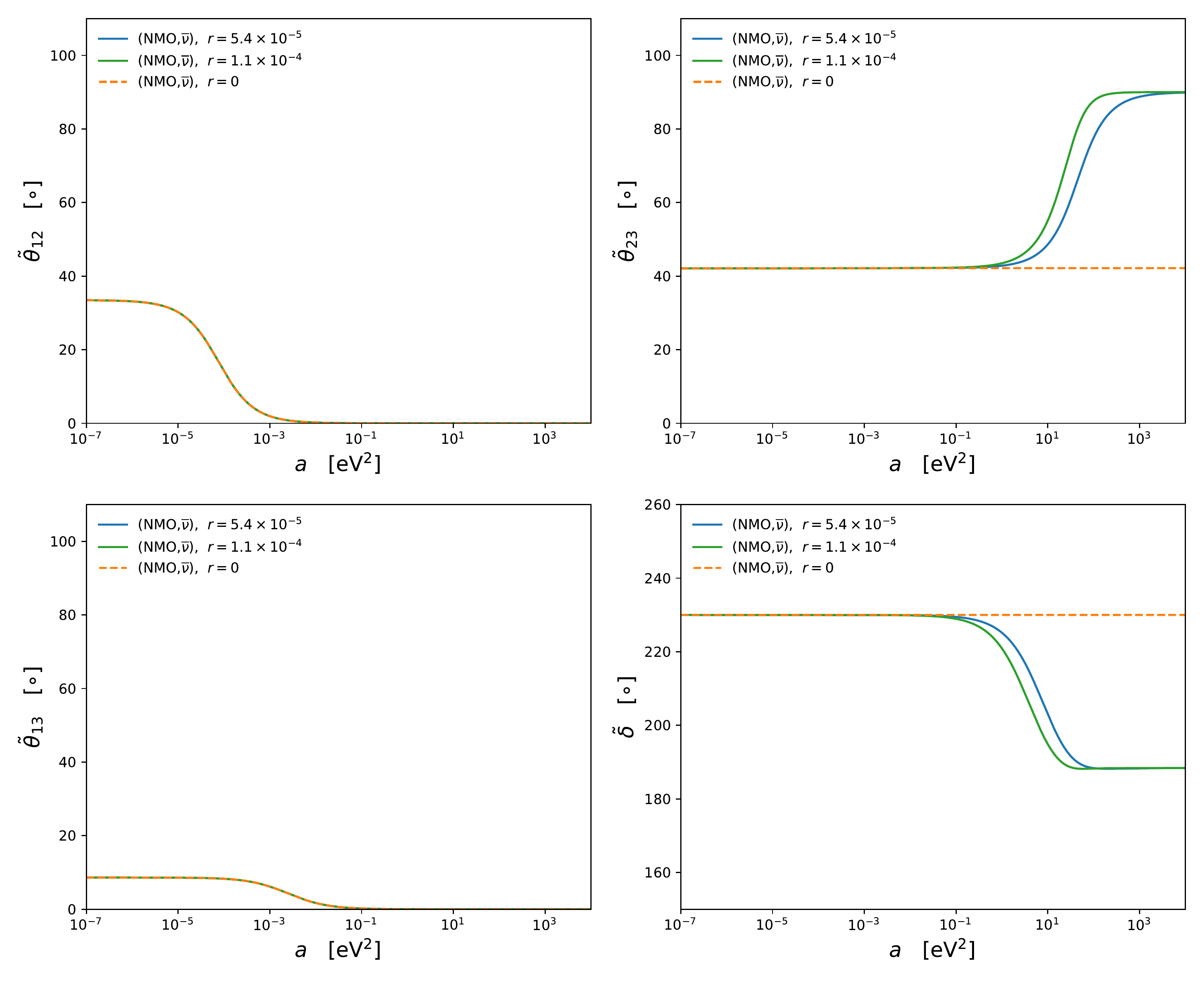}
\vspace{-0.6cm}
\caption{An illustration of the four effective flavor-mixing and CP-violating parameters
evolving with the matter parameter $a$ in the NMO case for antineutrino oscillations.}
\label{fig:NMO-antineutrino-angles}
\end{figure}
%%%%%%%%%%%%%%%%%%%%%%%%%%%%%%%%%%%%%%%%%%%%%%%%%%%%%%%%%%%%%%%%%%%%%%%%%%%%%%%%%%%%%%%%%%%%%
%%%%%%%%%%%%%%%%%%%%%%%%%%%%%%%%%%% Figure 6 %%%%%%%%%%%%%%%%%%%%%%%%%%%%%%%%%%%%%%%%%%%%%%%
\begin{figure}[!h]
\centering
\includegraphics[width=1\textwidth]{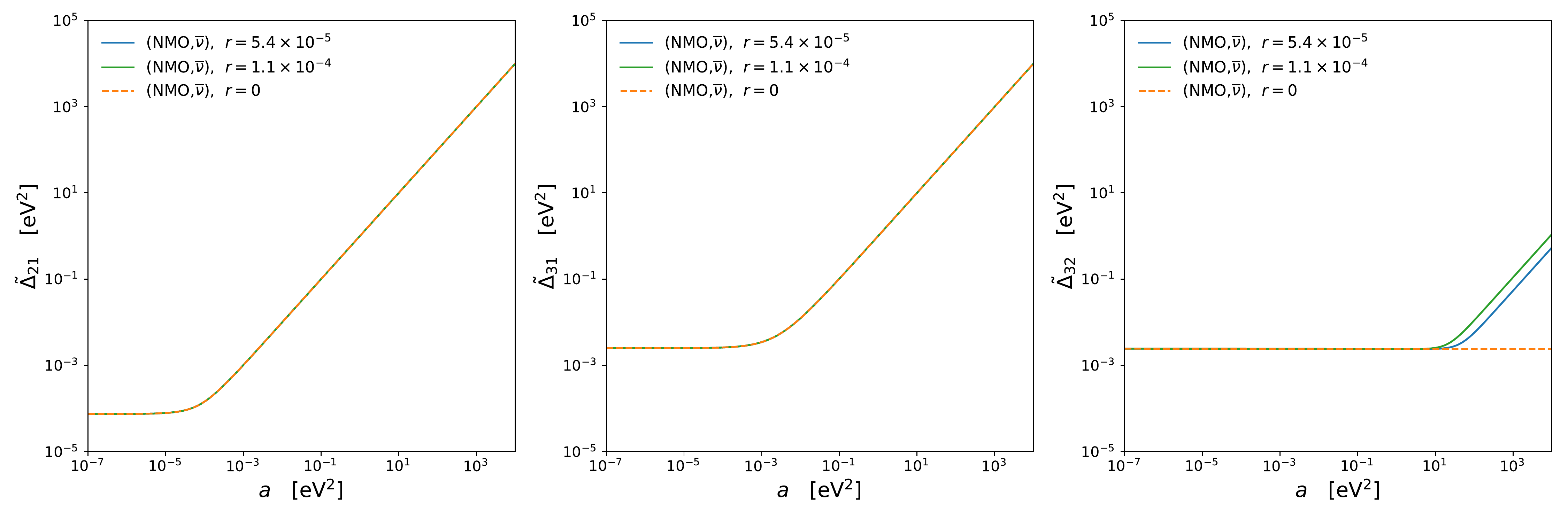}
\vspace{-1.1cm}
\caption{An illustration of the three effective neutrino mass-squared differences
evolving with the matter parameter $a$ in the NMO case for antineutrino oscillations.}
\label{fig:NMO-antineutrino-m2}
\end{figure}
%%%%%%%%%%%%%%%%%%%%%%%%%%%%%%%%%%%%%%%%%%%%%%%%%%%%%%%%%%%%%%%%%%%%%%%%%%%%%%%%%%%%%%%%%%%%

(3) {\it The NMO case for antineutrino oscillations} (Figs.~\ref{fig:NMO-antineutrino-angles}
and \ref{fig:NMO-antineutrino-m2}). As for an antineutrino beam travelling in
matter, the right-hand sides of Eqs.~(\ref{eq:RGE-mass-squared}) and (\ref{eq:angles}) need
to be multiplied by a negative sign, but Eq.~(\ref{eq:phase}) keeps unchanged. Note that $\widetilde{\Delta}_{31} \simeq \widetilde{\Delta}_{21}
\gg \widetilde{\Delta}_{32}$ when $a \gtrsim 0.1~{\rm eV}^2$ holds, in which case both
$\tilde{\theta}^{}_{12}$ and $\tilde{\theta}^{}_{13}$ approach zero as they are insensitive
to the one-loop radiative correction described by $r \neq 0$. In comparison,
$\tilde{\theta}^{}_{23}$ and $\tilde{\delta}$ are sensitive to this quantum effect for
$a \gtrsim 0.1~{\rm eV}^2$. The asymptotic value $\tilde{\theta}^{}_{23} \simeq 90^\circ$
in the $a \to \infty$ limit can therefore be understood from
\begin{eqnarray}
\dot{\tilde{\theta}}^{}_{23} \simeq  \frac{r \sin 2\tilde{\theta}^{}_{23}}
{2 \widetilde{\Delta}^{}_{32}} \to 0 \;
\label{eq:theta23NMO2}
%    (28)
\end{eqnarray}
which is quite similar to Eq.~(\ref{eq:theta23IMO}). Like the IMO case for neutrino
oscillations, here an analytical understanding of the asymptotic value
$\tilde{\delta} \simeq 188.4^{\circ}$ is not straightforward either.
%%%%%%%%%%%%%%%%%%%%%%%%%%%%%%%%%%% Figure 7 %%%%%%%%%%%%%%%%%%%%%%%%%%%%%%%%%%%%%%%%%%%%%%%
\begin{figure}[!h]
\centering
\includegraphics[width=0.9\textwidth]{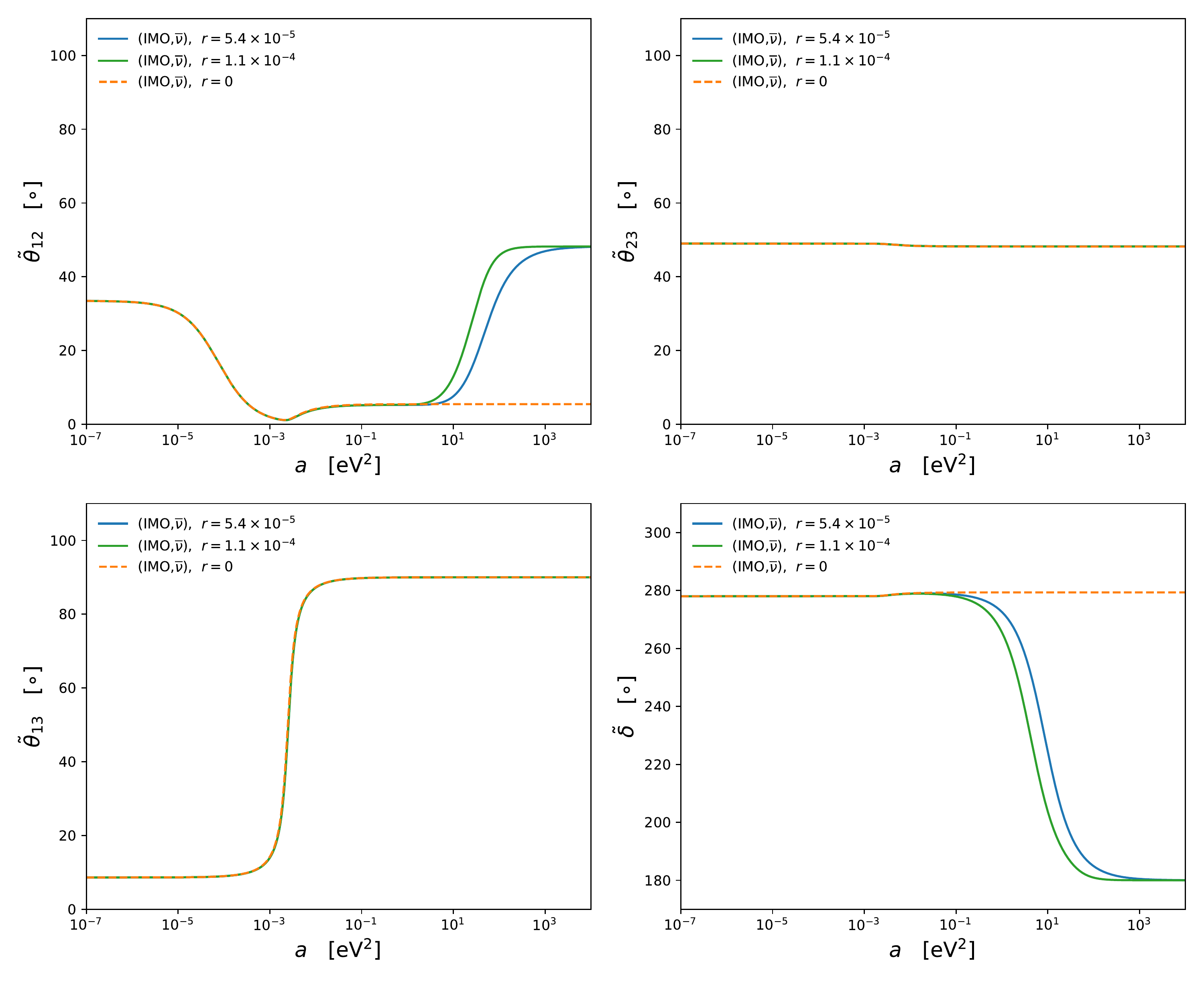}
\vspace{-0.6cm}
\caption{An illustration of the four effective flavor-mixing and CP-violating parameters
evolving with the matter parameter $a$ in the IMO case for antineutrino oscillations.}
\label{fig:IMO-antineutrino-angles}
\end{figure}
%%%%%%%%%%%%%%%%%%%%%%%%%%%%%%%%%%%%%%%%%%%%%%%%%%%%%%%%%%%%%%%%%%%%%%%%%%%%%%%%%%%%%%%%%%%%
%%%%%%%%%%%%%%%%%%%%%%%%%%%%%%%%%%% Figure 8 %%%%%%%%%%%%%%%%%%%%%%%%%%%%%%%%%%%%%%%%%%%%%%%
\begin{figure}[!h]
\centering
\includegraphics[width=1\textwidth]{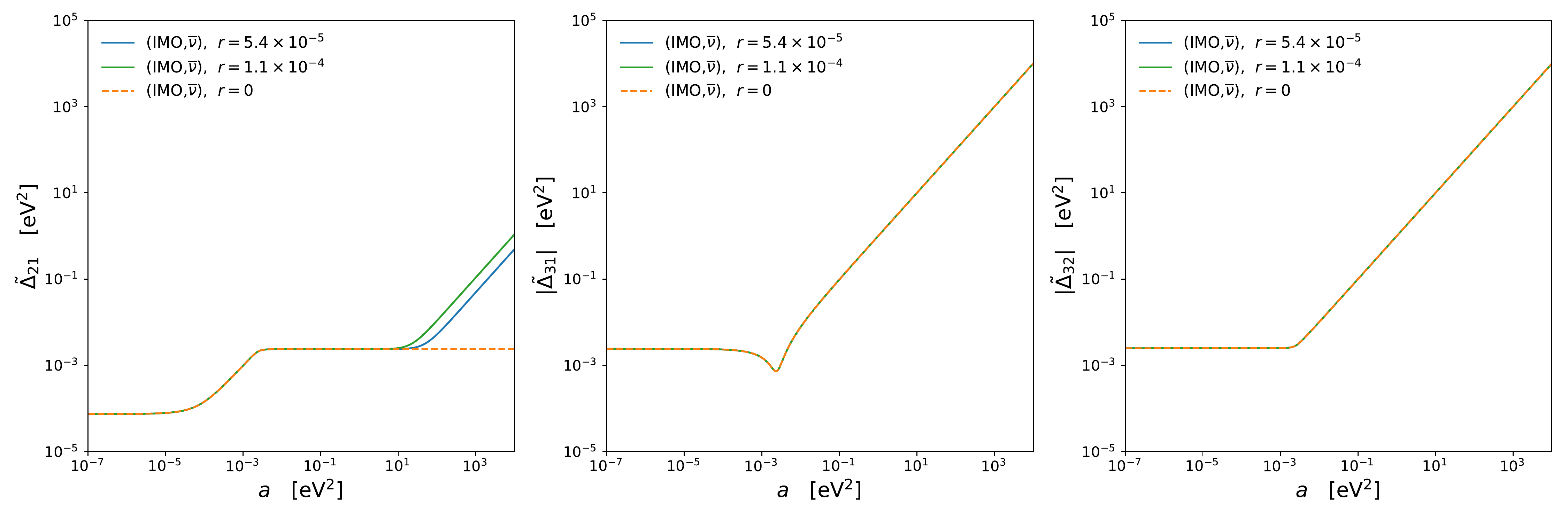}
\vspace{-1.1cm}
\caption{An illustration of the three effective neutrino mass-squared differences
evolving with the matter parameter $a$ in the IMO case for antineutrino oscillations.}
\label{fig:IMO-antineutrino-m2}
\end{figure}
%%%%%%%%%%%%%%%%%%%%%%%%%%%%%%%%%%%%%%%%%%%%%%%%%%%%%%%%%%%%%%%%%%%%%%%%%%%%%%%%%%%%%%%%%%%

(4) {\it The IMO case for antineutrino oscillations} (Figs.~\ref{fig:IMO-antineutrino-angles}
and \ref{fig:IMO-antineutrino-m2}). In this case we find that
$\big|\widetilde{\Delta}^{}_{32}\big|\simeq \big|\widetilde{\Delta}^{}_{31}\big|\gg
\widetilde{\Delta}_{21}$ holds for $a \gtrsim 0.1~{\rm eV}^2$, and the effective flavor mixing
angles $\tilde{\theta}^{}_{13}$ and $\tilde{\theta}^{}_{23}$ are insensitive to
the one-loop quantum effect because they depend only upon $r/ \widetilde{\Delta}^{}_{31}$ and
$r/ \widetilde{\Delta}^{}_{32}$ which are strongly suppressed in dense matter. In comparison,
$\tilde{\theta}^{}_{12}$ and $\tilde{\delta}$ become sensitive to $r \neq 0$ when
$a \gtrsim 0.1~{\rm eV}^2$ holds, and they approach their respective fixed points
$\tilde{\theta}^{}_{12} \simeq 48.2^\circ$ and $\tilde{\delta} \simeq 180^\circ$ for
$a \gtrsim 10^3~{\rm eV}^2$. To understand why $\tilde{\theta}^{}_{12}$ assumes such an
asymptotic value, let us take a look at the dominant term of $\dot{\tilde{\theta}}^{}_{12}$
in Eq.~(\ref{eq:angles}) by multiplying its right-hand side with a negative sign and
inputting $\tilde{\theta}^{}_{13} \simeq 90^\circ$ and
$\tilde{\delta} \simeq 180^\circ$ in the $a \to \infty$ limit,
\begin{eqnarray}
\dot{\tilde{\theta}}^{}_{12} \simeq
-\frac{r \sin 2\big(\tilde{\theta}^{}_{12} - \tilde{\theta}^{}_{23}\big)}
{2 \widetilde{\Delta}^{}_{21}} \to 0 \; .
\label{eq:theta12IMO2}
%    (29)
\end{eqnarray}
So we arrive at $\tilde{\theta}^{}_{12} \simeq \tilde{\theta}^{}_{23} \simeq 48.2^\circ$ at
the fixed points, where $\tilde{\theta}^{}_{23} \simeq 48.2^\circ$ can be
obtained from Eq.~(\ref{eq:theta23}) as a good
approximation~\cite{Freund:2001pn,Xing:2018lob,Cervera:2000kp}. Of course, these two
asymptotic values can also be read off directly from our numerical results shown in Fig.~\ref{fig:IMO-antineutrino-angles}.
%%%%%%%%%%%%%%%%%%%%%%%%%%%%%%%%%%%%% Figure 9 %%%%%%%%%%%%%%%%%%%%%%%%%%%%%%%%%%%%%%%%%%%
\begin{figure}[!t]
\centering
\subfigure{
\begin{minipage}[t]{0.47\textwidth}
\centering
\includegraphics[width=0.9\textwidth]{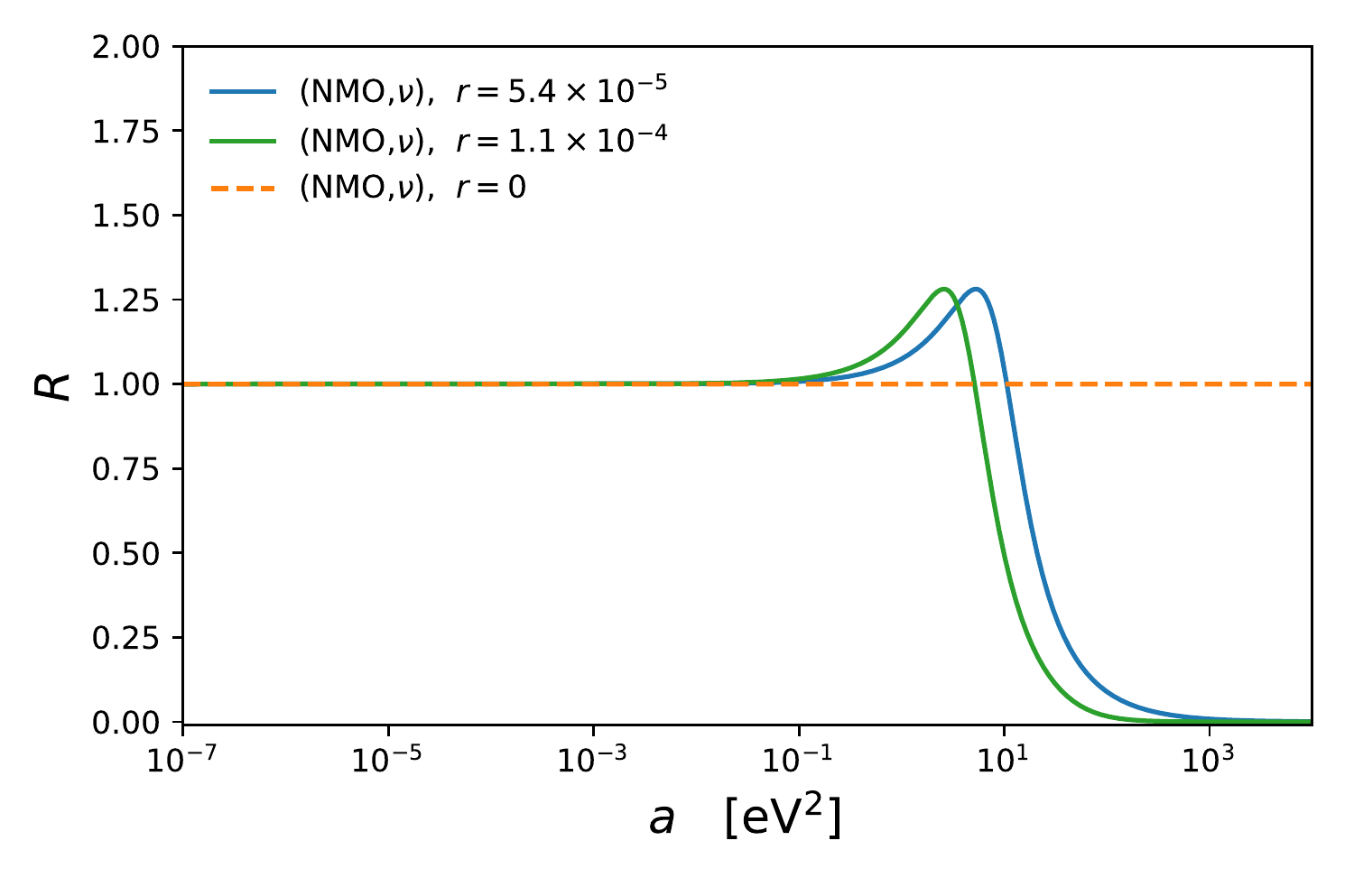}
\end{minipage}
}
\subfigure{
\begin{minipage}[t]{0.47\textwidth}
\centering
\includegraphics[width=0.9\textwidth]{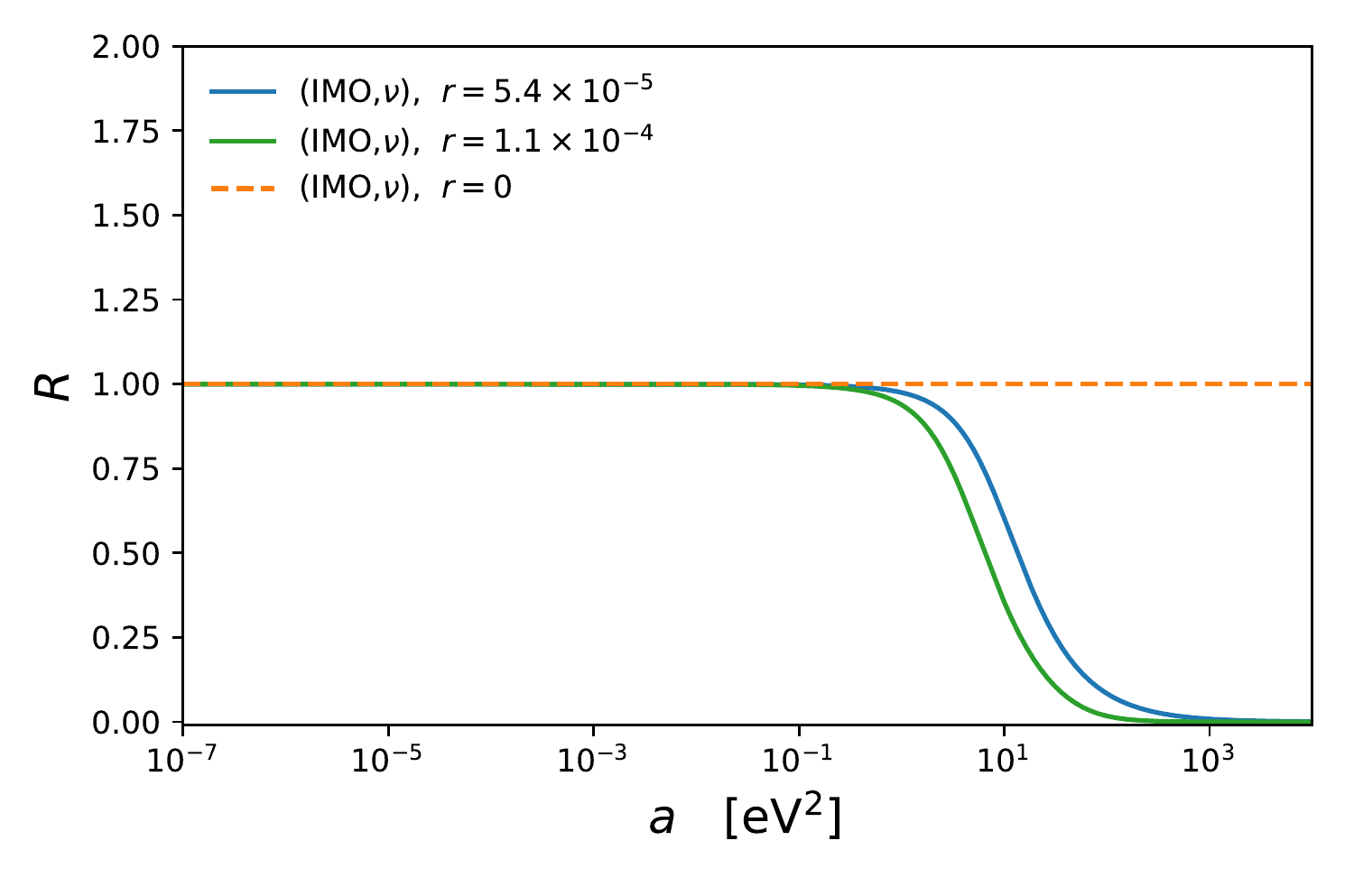}
\end{minipage}
}
\subfigure{
\begin{minipage}[t]{0.47\textwidth}
\centering
\includegraphics[width=0.9\textwidth]{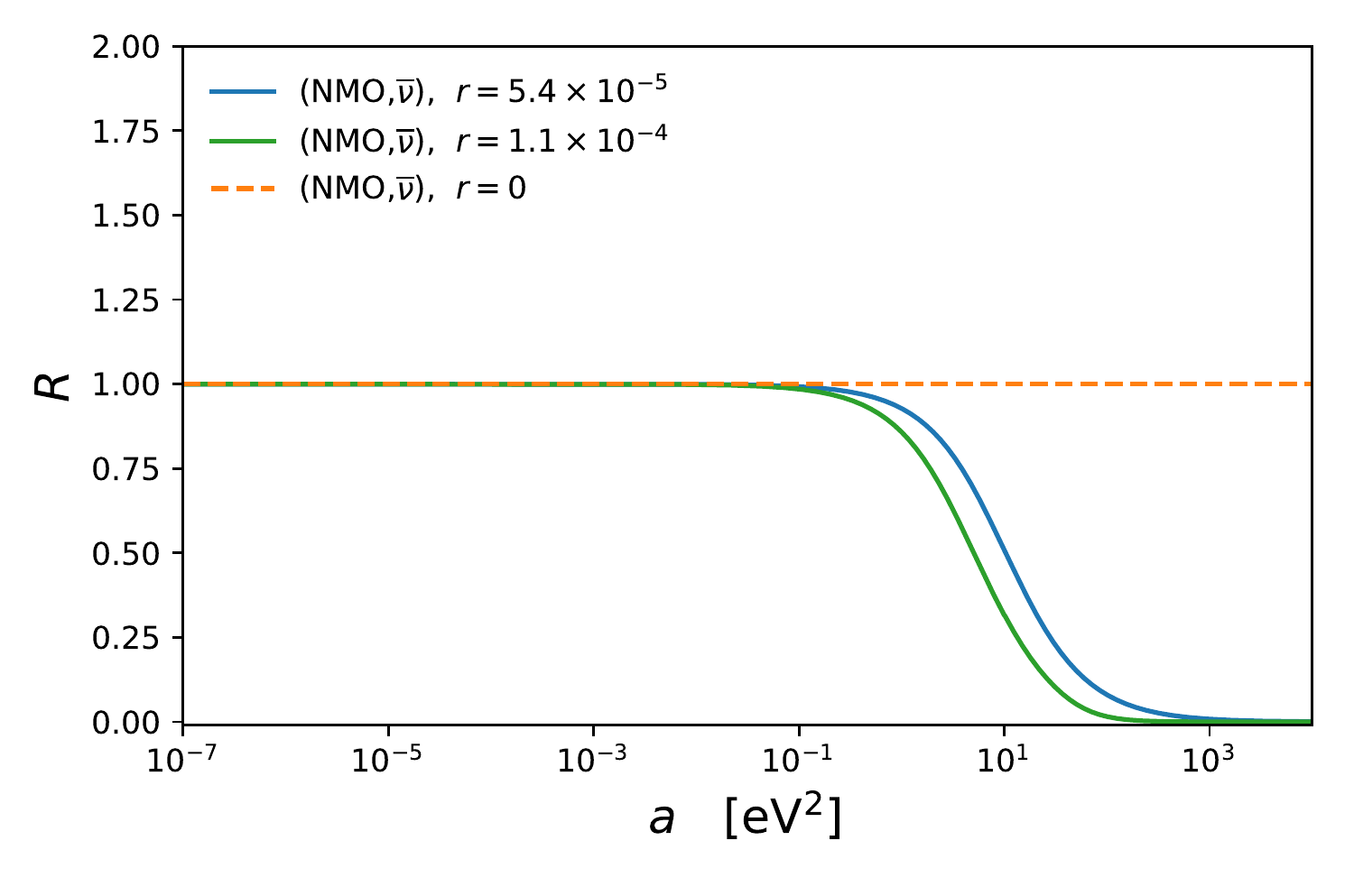}
\end{minipage}
}
\subfigure{
\begin{minipage}[t]{0.47\textwidth}
\centering
\includegraphics[width=0.9\textwidth]{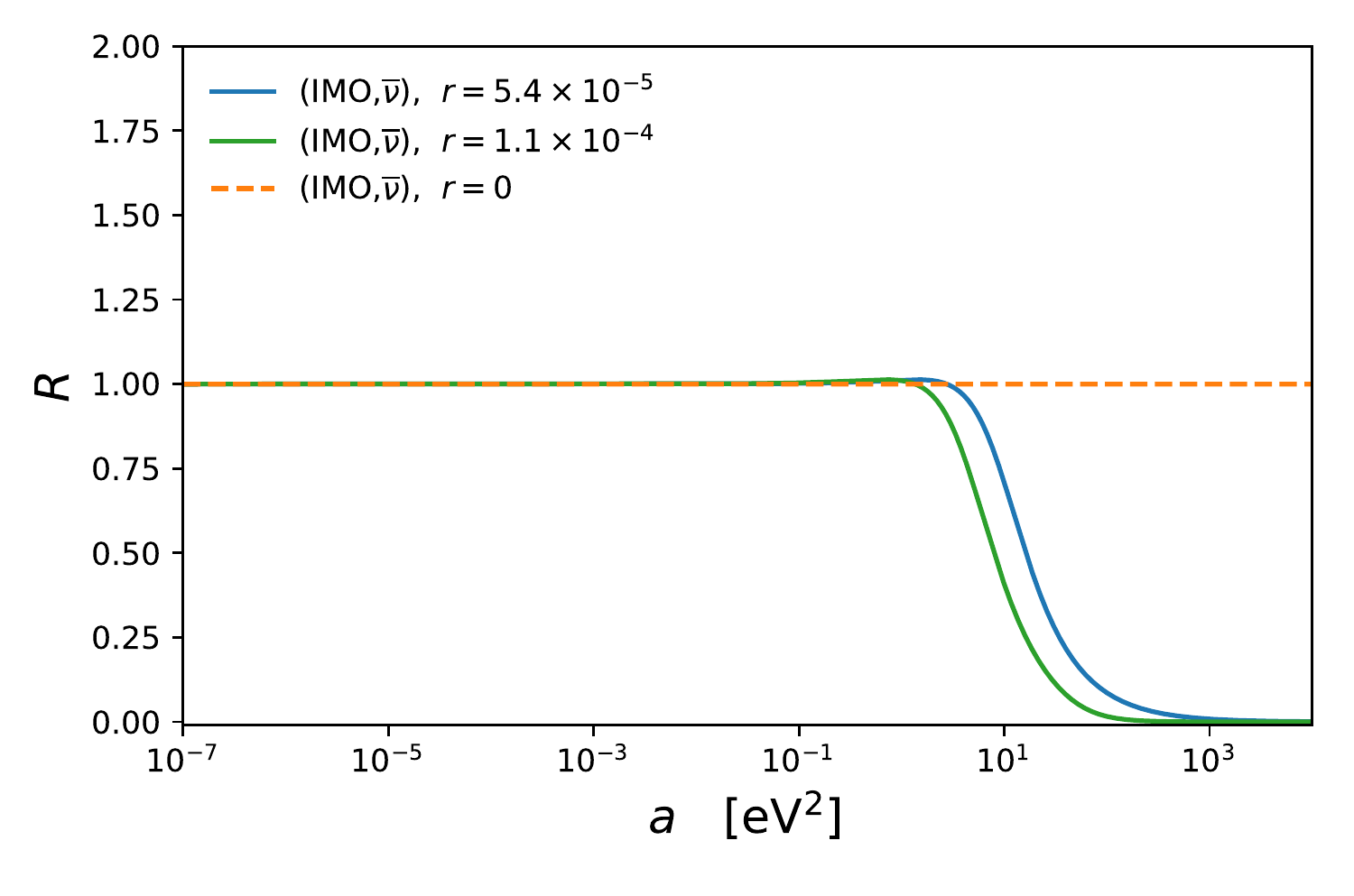}
\end{minipage}
}
\vspace{-0.6cm}
\caption{An illustration of the one-loop radiative correction to the tree-level Toshev relation
in dense matter, where $R \equiv \big(\sin 2\tilde{\theta}^{}_{23} \sin\tilde{\delta}\big)/
\big(\sin 2\theta^{}_{23} \sin\delta\big)$ has been defined in Eq.~(\ref{eq:Toshev-one-loop})
and its deviation from one signifies the quantum effect.}
\label{fig:Toshev relation}
\end{figure}
%%%%%%%%%%%%%%%%%%%%%%%%%%%%%%%%%%%%%%%%%%%%%%%%%%%%%%%%%%%%%%%%%%%%%%%%%%%%%%%%%%%%%%%%%%%%%
%%%%%%%%%%%%%%%%%%%%%%%%%%%%%%%%%%%%% Figure 10 %%%%%%%%%%%%%%%%%%%%%%%%%%%%%%%%%%%%%%%%%%%
\begin{figure}[!h]
\centering
\subfigure{
\begin{minipage}[t]{0.47\textwidth}
\centering
\includegraphics[width=0.9\textwidth]{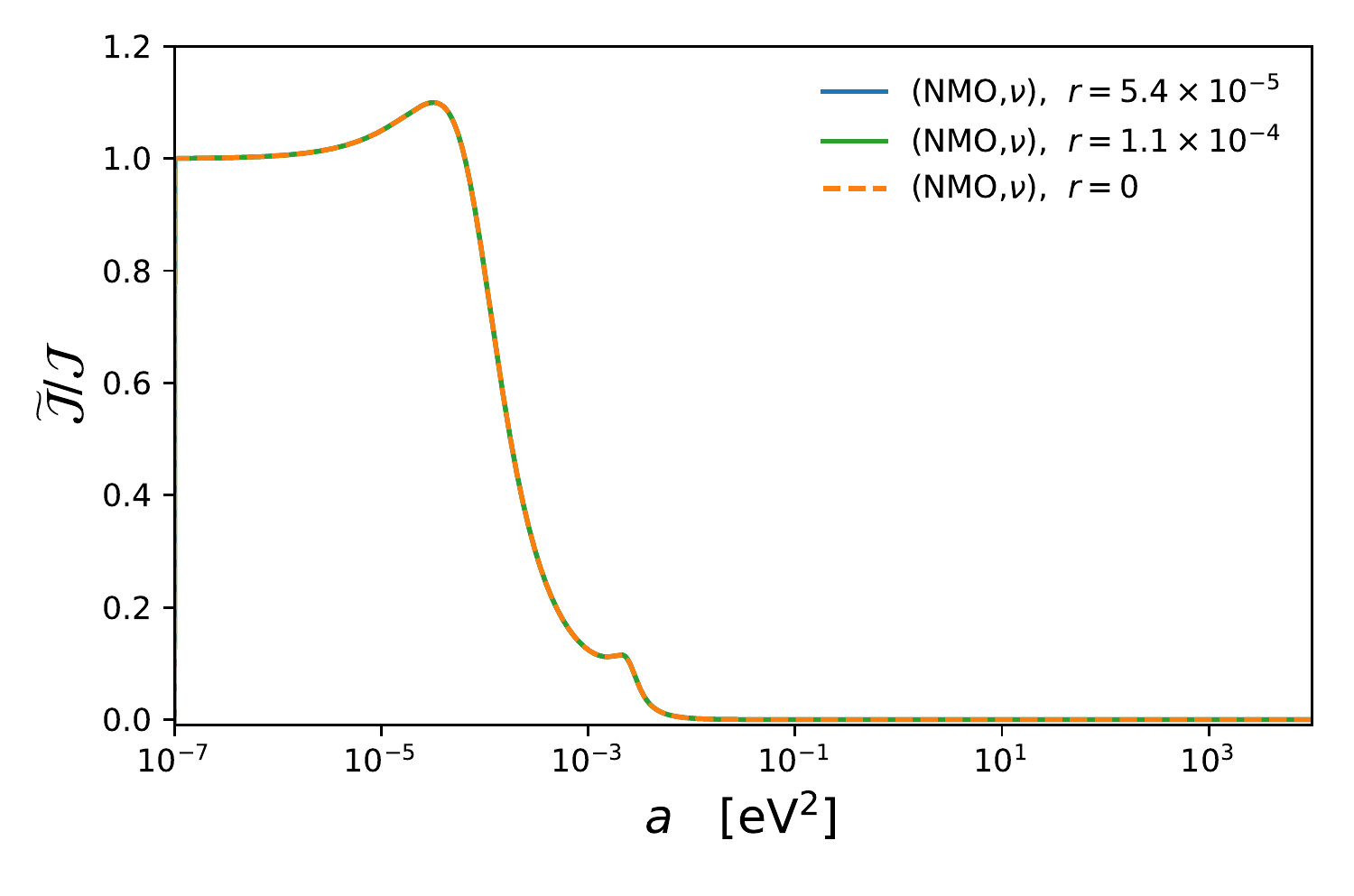}
\end{minipage}
}
\subfigure{
\begin{minipage}[t]{0.47\textwidth}
\centering
\includegraphics[width=0.9\textwidth]{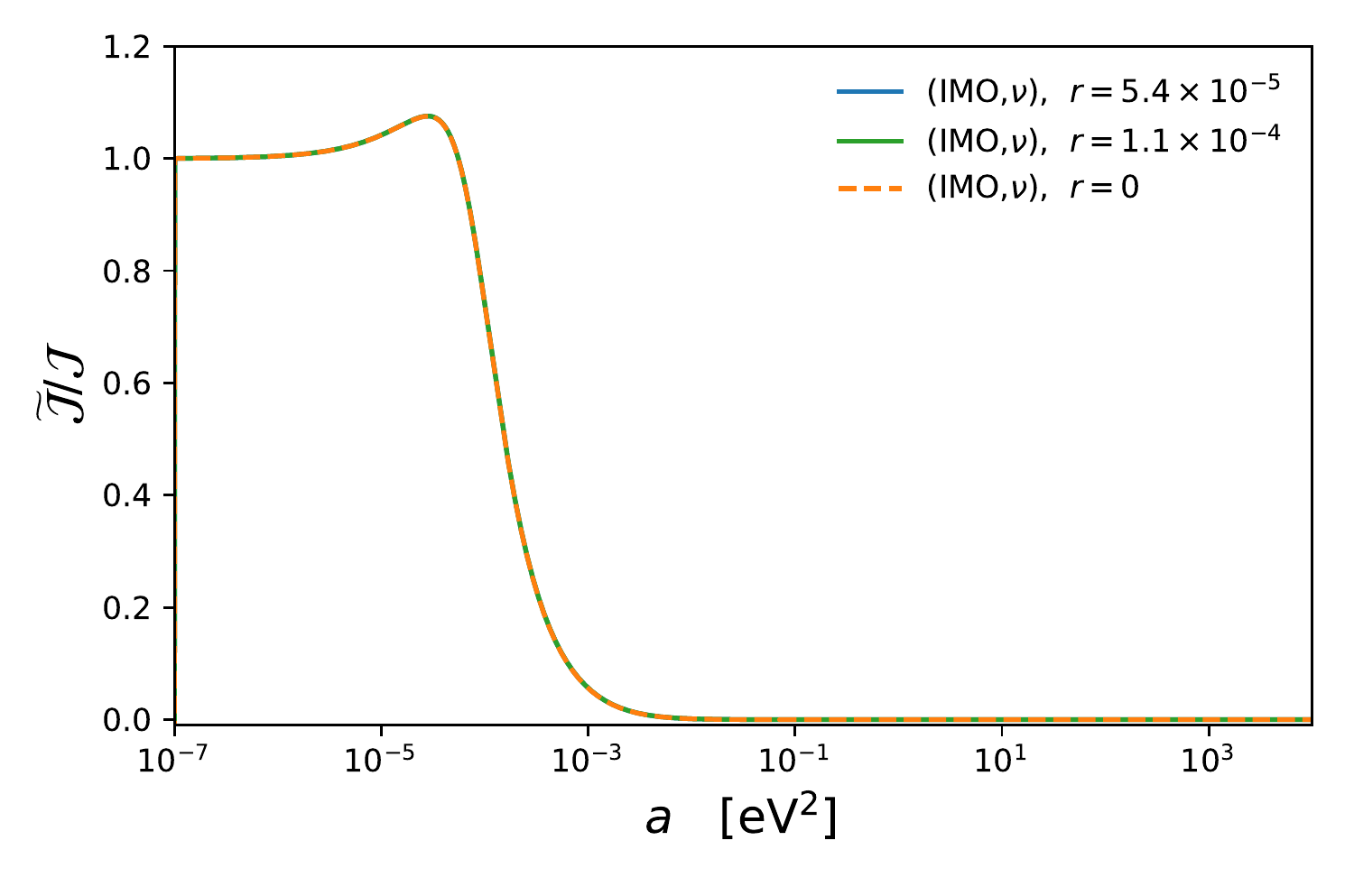}
\end{minipage}
}
\subfigure{
\begin{minipage}[t]{0.47\textwidth}
\centering
\includegraphics[width=0.9\textwidth]{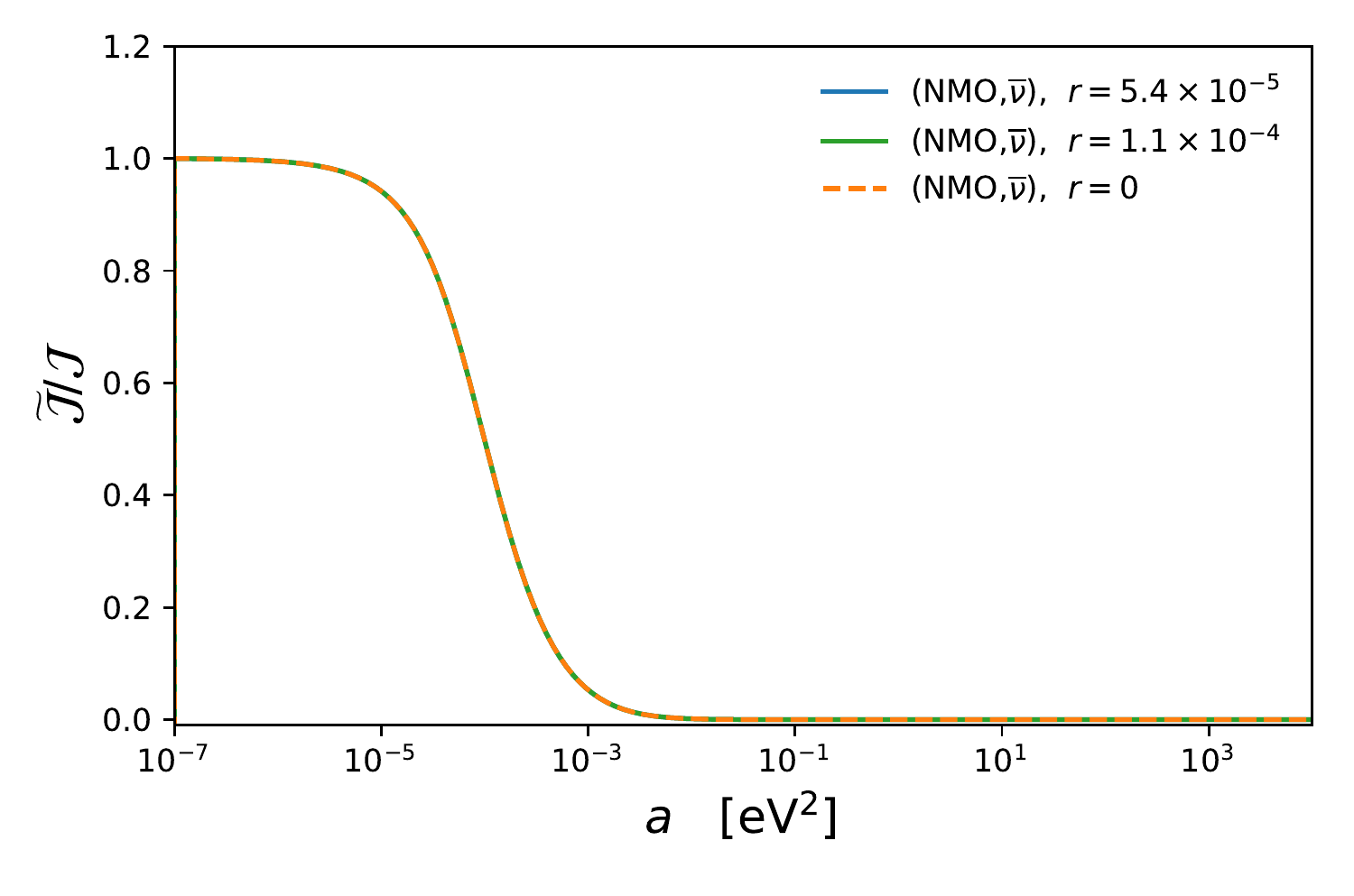}
\end{minipage}
}
\subfigure{
\begin{minipage}[t]{0.47\textwidth}
\centering
\includegraphics[width=0.9\textwidth]{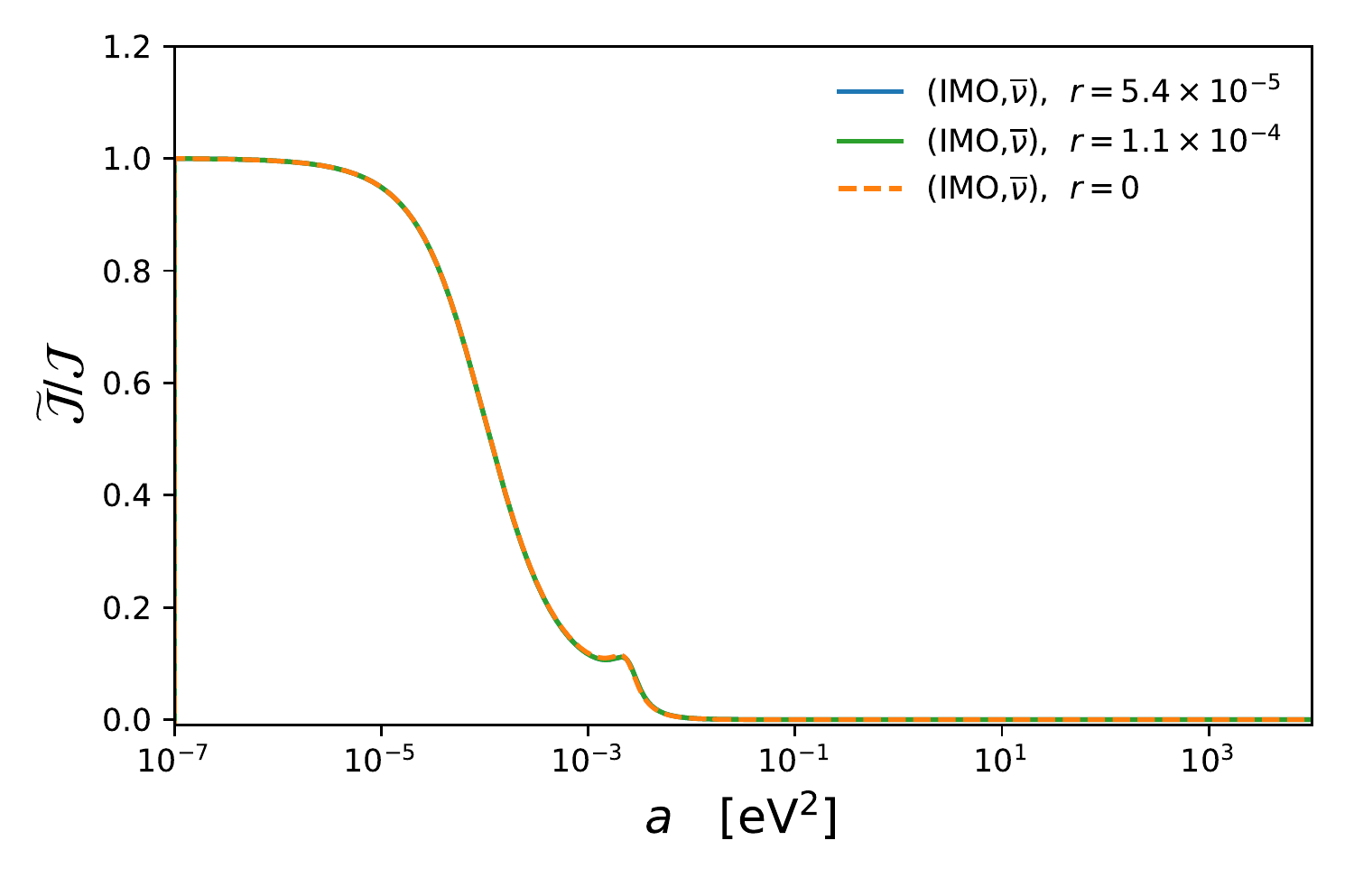}
\end{minipage}
}
\vspace{-0.6cm}
\caption{An illustration of the one-loop radiative correction to the effective Jarlskog 
invariant $\tilde{J}$, which is normalized by its fundamental counterpart $J$,
in dense matter.}
\label{fig:Jarlskog invariant}
\end{figure}
%%%%%%%%%%%%%%%%%%%%%%%%%%%%%%%%%%%%%%%%%%%%%%%%%%%%%%%%%%%%%%%%%%%%%%%%%%%%%%%%%%%%%%%%%%%%%

To numerically illustrate to what extent the tree-level Toshev relation can be
modified by the one-loop radiative correction in a dense medium with either
$r \simeq 5.4 \times 10^{-5}$ (i.e., $N^{}_n = N^{}_p = N^{}_e$) or
$r \simeq 1.1 \times 10^{-4}$ (i.e., $N^{}_n = 3N^{}_p = 3N^{}_e$),
we plot the evolution of $R$ defined by Eq.~(\ref{eq:Toshev-one-loop}) with
respect to the matter parameter $a$ in Fig.~\ref{fig:Toshev relation} for both
neutrino and antineutrino oscillations
in both the NMO and IMO cases. It is clear that $R \simeq 1$ is an excellent
approximation for $a \lesssim 0.1~{\rm eV}^2$, but the quantum effect becomes significant
as $a$ is much larger. In particular, $R \to 0$ when $a \gtrsim 10^3~{\rm eV}^2$ holds for
all the four cases, as a straightforward consequence of either $\tilde{\delta} \to 0^\circ$
(or $180^\circ$) or $\tilde{\theta}^{}_{23} \to 90^\circ$. Note that there is a peak in the
curve of $R$ when $a$ is close to a few ${\rm eV}^2$ and up to about $10~{\rm eV}^2$ in the
NMO case for neutrino oscillations or in the IMO case for antineutrino oscillations.
In either of these two scenarios $\tilde{\theta}^{}_{23}$ almost keeps unchanged
but $\tilde{\delta}$ crosses its threshold value $\tilde{\delta} = 270^\circ$ as can
be seen in Fig.~\ref{fig:NMO-neutrino-angles} or Fig.~\ref{fig:IMO-antineutrino-angles},
which allows $R \propto \sin\tilde{\delta} = -1$ to have a local peak.

As an interesting by-product, the ratio of the effective Jarlskog invariant $\tilde{J}$ in 
matter to the fundamental Jarlskog invariant $J$ in vacuum is numerically calculated and its
evolution with the matter parameter $a$ is illustrated in Fig.~\ref{fig:Jarlskog invariant}
for both the NMO and IMO cases and for both neutrino and antineutrino oscillations. Now
that the Naumov relation in Eq.~(\ref{eq:Naumov}) keeps valid at the one-loop level, our
result is certainly insensitive to $r \neq 0$ and thus consistent with those obtained 
previously at the tree level (see, e.g., Refs.~\cite{Xing:2018lob,Xing:2016ymg}). 
Note that one of the three effective neutrino mass-squared differences 
$\widetilde{\Delta}^{}_{ji}$ (for $ji = 21, 31, 32$) is actually
sensitive to $r \neq 0$ when $a \gtrsim 10~{\rm eV}^2$ holds, as one can see from 
Figs.~\ref{fig:NMO-neutrino-m2}, \ref{fig:IMO-neutrino-m2}, \ref{fig:NMO-antineutrino-m2}
and \ref{fig:IMO-antineutrino-m2}. But this effect does not appreciably manifest itself 
in Fig.~\ref{fig:Jarlskog invariant}, simply because $\tilde{J}$ has already approaches
zero for $a \gtrsim 0.1~{\rm eV}^2$. 

\section{Summary}

Although the one-loop electroweak radiative corrections to coherent forward neutrino
scattering in a medium may slightly break the tree-level universality of
neutral-current contributions of three active neutrino flavors to the matter potential
term of the effective Hamiltonian that is responsible for neutrino oscillations in matter,
they have not attracted much attention because their effects on those currently available
experiments of solar, atmospheric, reactor and accelerator neutrino oscillations are
negligibly small. However, such quantum corrections are expected to be significant when
studying neutrino oscillations in dense matter. In this paper we have examined this kind
of small but nontrivial quantum effect by deriving the RGE-like differential equations
for the relevant effective neutrino oscillation quantities with respect to the matter
parameter $a$. The tree-level Toshev relation $\sin 2 \tilde{\theta}^{}_{23}
\sin\tilde{\delta} = \sin 2\theta^{}_{23} \sin\delta$, which links the genuine
flavor-mixing and CP-violating parameters $\big(\theta^{}_{23}, \delta\big)$ in vacuum
to their effective counterparts $\big(\tilde{\theta}^{}_{23}, \tilde{\delta}\big)$
in matter, is found to be modified at the one-loop level. In comparison, we have
shown that the tree-level Naumov relation $\tilde{J} \widetilde{\Delta}^{}_{21}
\widetilde{\Delta}^{}_{31} \widetilde{\Delta}^{}_{32} = J \Delta^{}_{21} \Delta^{}_{31} 
\Delta^{}_{32}$ keeps valid at the one-loop level. A numerical illustration of
the significance of such a one-loop quantum effect has also been presented for neutrino
and antineutrino oscillations in dense matter.

We emphasize that the Toshev relation may provide a simple but instructive way to
test matter effects on flavor mixing and CP violation in neutrino (or antineutrino)
oscillations in the upcoming precision measurement era, especially when both the
neutrino and antineutrino beams can be prepared and studied in a single experiment. 

It is no doubt that our analysis made above is conceptually important and may even find 
some useful applications in exploring the phenomena of neutrino and antineutrino 
oscillations in a dense matter environment. On the other hand, this work provides a new 
example which supports the second Weinberg's law of progress in theoretical physics, 
namely ``{\it Do not trust arguments based on the lowest order of perturbation 
theory}"~\cite{Weinberg:1981qq}.

\vspace{0.5cm}

\textsl{One of us (Z.Z.X.) is indebted to Shun Zhou for useful discussions.
This work is supported by the National Natural Science Foundation of China under
grants No. 12075254 and No. 11835013.}

\end{document}